\documentclass[twocolumn,aps,pra,showpacs]{revtex4}
 \pdfoutput=1
\usepackage[T1]{fontenc}
\usepackage[latin9]{inputenc}
\usepackage{amsmath}
\usepackage{amssymb}
\usepackage{graphicx}
\usepackage{esint}

\makeatletter

\newcommand*\LyXThinSpace{\,\hspace{0pt}}

\@ifundefined{textcolor}{}
{%
 \definecolor{BLACK}{gray}{0}
 \definecolor{WHITE}{gray}{1}
 \definecolor{RED}{rgb}{1,0,0}
 \definecolor{GREEN}{rgb}{0,1,0}
 \definecolor{BLUE}{rgb}{0,0,1}
 \definecolor{CYAN}{cmyk}{1,0,0,0}
 \definecolor{MAGENTA}{cmyk}{0,1,0,0}
 \definecolor{YELLOW}{cmyk}{0,0,1,0}
}

\makeatother

\begin{document}

\title{Three-dimensional spin-orbit coupled Fermi gases: Fulde-Ferrell pairing,
Majorana fermions, Weyl fermions and gapless topological superfluidity}

\author{Xia-Ji Liu$^{1}$, Hui Hu$^{1}$, and Han Pu$^{2}$ }

\email{hpu@rice.edu}

\affiliation{$^{1}$Centre for Quantum and Optical Science, Swinburne University
of Technology, Melbourne 3122, Australia}

\affiliation{$^{2}$Department of Physics and Astronomy, and Rice Quantum Institute,
Rice University, Houston, TX 77251, USA}

\date{\today}
\begin{abstract}
We theoretically investigate a three-dimensional Fermi gas with Rashba
spin-orbit coupling in the presence of both out-of-plane and in-plane
Zeeman fields. We show that, driven by a sufficiently large Zeeman
field, either out-of-plane or in-plane, the superfluid phase of this
system exhibits a number of interesting features, including inhomogeneous
Fulde-Ferrell pairing, gapped or gapless topological order and exotic
quasi-particle excitations known as Weyl fermions that have linear
energy dispersions in momentum space (i.e., massless Dirac fermions).
The topological superfluid phase can have either four or two topologically
protected Weyl nodes. We present the phase diagrams at both zero and
finite temperatures and discuss the possibility of their observation
in an atomic Fermi gas with synthetic spin-orbit coupling. In this
context, topological superfluid phases with an imperfect Rashba spin-orbit
coupling are also studied.
\end{abstract}


\maketitle

\section{Introduction}

Quantum simulations of intriguing many-body problems with ultracold
atoms have now become a paradigm in different fields of physics \cite{Bloch2012}.
The major advantage of ultracold atoms is their unprecedented controllability
in tuning interactions, dimensionality, populations and species of
atoms, which constitutes an ideal toolbox for understanding the consequence
of strong interactions \cite{Chin2010}. Most recently, a new tool
- a synthetic non-Abelian gauge field or spin-orbit coupling - was
added to the toolbox \cite{Lin2011}. In condensed matter physics,
it is well known that spin-orbit coupling plays a key role in new-generation
solid-state materials such as topological insulators \cite{Hasan2010,Qi2011},
quantum spin Hall systems \cite{Xiao2010} and non-centrosymmetric
superconductors \cite{Yip2014}. In this respect, recent realizations
of synthetic spin-orbit coupling in ultracold atoms open an entirely
new direction to quantum simulate and characterize these new-generation
materials, without the need to overcome the intrinsic complexity in
compositions and interactions, which is often encountered in solid-state
systems. 

\begin{figure}
\begin{centering}
\includegraphics[clip,width=0.48\textwidth]{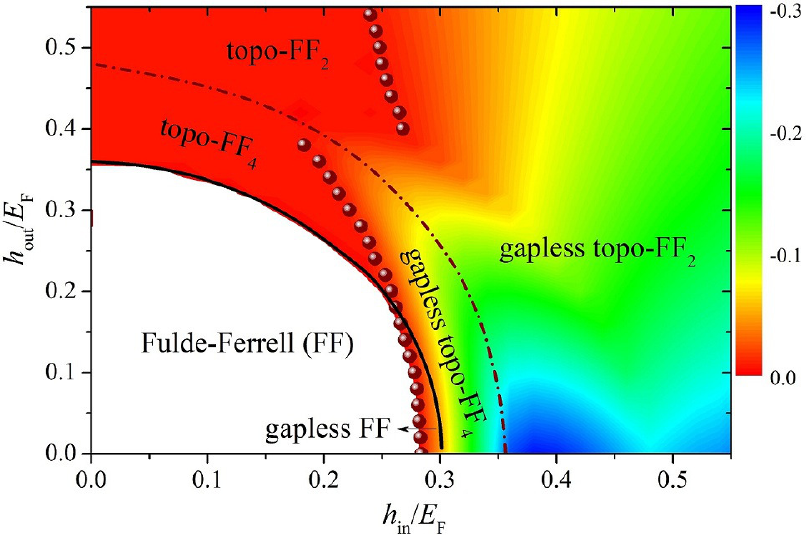} 
\par\end{centering}

\protect\caption{(Color online) Zero temperature phase diagram of a 3D Rashba spin-orbit
coupled Fermi gas on the plane of in-plane and out-of-plane Zeeman
fields, with a spin-orbit coupling strength $\lambda=E_{F}/k_{F}$
and an interaction strength $1/(k_{F}a_{s})=-0.5$. The color map
shows the lowest energy in the lower particle branch, in units of
$E_{F}$. There are various superfluid phases, as described in detail
in the text. The solid line show the topological phase transition.
While the dot-dashed line indicates the transition within the two
types of topological superfluids, which have four or two Weyl nodes/loops,
respectively. The solid circles are the boundary where superfluid
becomes gapless, apart from the Weyl nodes. }

\label{fig1} 
\end{figure}

Indeed, over the past few years, there have been a tremendous amount
of theoretical works along this direction \cite{reviewSOC_ZLHP,reviewSOC_Zhai,reviewSOC_Xu,Vyasanakere2011,Zhu2011,Hu2011,Yu2011,Gong2011,Jiang2011,Liu2012a,Zhou2012,He2012,Anderson2012,Gong2012,Seo2012,Liu2012b,Wei2012,Hu2013PRL,He2013,Zheng2013,Dong2013PRA,Wu2013,Liu2013a,Dong2013NJP,Zhou2013,Hu2013NJP,Seo2013,Iskin2013,Anderson2013,Chen2013,Liu2013b,Qu2013,Zhang2013,Shenoy2013,Liu2013c,Hu2014,Xu2014,Cao2014a,Devreese2014,Jiang2014,Zheng2014,Cao2014b},
triggered by the demonstration of a particular spin-orbit coupling
in ultracold atomic Fermi gases by the two-photon Raman technique,
which has an equal-weight superposition of Rashba and Dresselhaus
spin-orbit couplings \cite{Wang2012,Cheuk2012,Williams2013,Fu2013}.
At the early stage, the possibility of observing a peculiar two-body
bound state and the related anisotropic superfluid induced by Rashba
spin-orbit coupling was discussed \cite{Vyasanakere2011,Hu2011,Yu2011}.
In the presence of out-of-plane Zeeman field, the routine to a topological
superfluid and Majorana fermions in two dimensions \cite{Zhu2011,Liu2012a,Gong2012,He2013,Wilczek2009}
or to a Weyl superfluid in three dimensions \cite{Gong2011,Seo2013,Sau2012}
- which has exotic quasiparticles with linear energy dispersion in
momentum space - was addressed. Later on, it was realized that unconventional
Fulde-Ferrell pairing could be significantly enhanced by spin-orbit
coupling in combination with in-plane Zeeman field due to the distorted
Fermi surface \cite{Zheng2013,Dong2013PRA,Wu2013,Liu2013a,Dong2013NJP,Zhou2013,Hu2013NJP,Iskin2013,Shenoy2013,Liu2013c}.
As a result, spin-orbit coupled Fermi gases provide an ideal candidate
to investigate the long-sought Fulde-Ferrell superfluidity or superconductivity
\cite{Fulde1964}, which is yet to be unambiguously confirmed in solid-state
systems. Very recently, it is was shown that the Fulde-Ferrell pairing
may coexist with the topological order, leading to the predictions
of topological Fulde-Ferrell superfluids \cite{Chen2013,Liu2013b,Qu2013,Zhang2013,Jiang2014},
including a brand-new quantum matter of \emph{gapless} topological
superfluid that has no analogies in solid-state systems \cite{Hu2014,Cao2014a,Cao2014b}.

In this work, by using a spin-orbit coupled Fermi gas in three dimensions
with both in-plane and out-of-plane Zeeman fields as a prototype,
we show that the above-mentioned concepts in unconventional superfluid
phases, including Fulde-Ferrell pairings, Majorana fermions, Weyl
fermions and topological superfluidity, can be examined and demonstrated
in near-future experiments. Our main result is summarized in Fig.
\ref{fig1}, which displays a very rich phase diagram at zero temperature
and at an interaction parameter $1/(k_{F}a_{s})=-0.5$ and a spin-orbit
coupling strength $\lambda=E_{F}/k_{F}$, where $k_{F}$ and $E_{F}=\hbar^{2}k_{F}^{2}/(2m)$
are respectively the Fermi wavevector and Fermi energy, and $a_{s}$
is the $s$-wave scattering length. By tuning the in-plane and out-of-plane
Zeeman fields, the Fermi system can support various intriguing unconventional
superfluid phases: the Fulde-Ferrell superfluid (FF), topologically
non-trivial Fulde-Ferrell superfluid with four (topo-FF$_{4}$) or
two Weyl nodes (topo-FF$_{2}$), gapless Fulde-Ferrell superfluid
(gapless FF), and gapless topologically non-trivial Fulde-Ferrell
superfluid with four (gapless topo-FF$_{4}$) or two Weyl nodes/loops
(gapless topo-FF$_{2}$). The purpose of this work is to discuss in
detail the emergence of these interesting superfluid phases, their
thermodynamic stability and their dependence on the form of spin-orbit
coupling.

We note that the same Fermi system has been considered earlier by
Xu \textit{et al.} with the focus on the anisotropic Weyl fermions
due to an in-plane Zeeman field \cite{Xu2014}. Our results differ
theirs in several ways: (1) We clarify that the in-plane Zeeman field
itself can drive a topological phase transition and lead to a gapless
topological superfluid; (2) In the topological phases, a Fulde-Ferrell
superfluid, either gapped or gapless, can be further classified according
to the number of Weyl nodes (in the gapped area) or Weyl loops (in
the gapless region). With increasing Zeeman field, a topologically
trivial phase first becomes a topologically non-trivial phase with
four Weyl nodes/loops and then turns into the one with two Weyl nodes/loops
when the Zeeman field is sufficiently large; Majorana surface states
in different superfluid phases are discussed. (3) Various finite-temperature
phase diagrams are presented, which are useful for future experiments
that have to be carried out at non-zero temperatures; (4) We consider
alternative form of spin-orbit coupling away from the Rashba limit
and show different intriguing topological phases do not depend critically
on the type of spin-orbit coupling. 

On the other hand, we also note that our theoretical investigations
are based on mean-field theory. The beyond mean-field effect has been
recently considered by Zheng and co-workers \cite{Zheng2014}, by
using a pseudogap approximation for pair fluctuations.

Our paper is organized as follows. In the next section, we introduce
the mean-field theoretical framework. In Sec. III, we discuss in detail
the conditions for different unconventional superfluid phases, their
energy spectrum and the surface states when a hard-wall potential
is imposed at the (opposite) two boundaries. Finite temperature phase
diagrams at different Zeeman fields are constructed. We also investigate
how the zero-temperature phase diagram is affected by the form of
spin-orbit coupling. Finally, Sec. IV is devoted to the conclusions
and outlooks.

\section{Theoretical framework}

We consider a three-dimensional (3D) spin-1/2 Fermi gas of $^{6}$Li
or $^{40}$K atoms near broad Feshbach resonances with two-dimensional
(2D) spin-orbit coupling $\lambda_{z}\hat{k}_{z}\sigma_{z}+\lambda_{x}\hat{k}_{x}\sigma_{x}$
in the presence of an in-plane Zeeman field $h_{\textrm{in}}\sigma_{z}$
along the $z$-direction (i.e., on the $z$-$x$ plane of spin-orbit
coupling) and an out-of-plane Zeeman field $h_{\textrm{out}}\sigma_{y}$
along the $y$-direction perpendicular to the plane of spin-orbit
coupling \cite{Hu2013NJP,Xu2014}. Here, $\hat{k}_{x}\equiv-i\partial_{x}$,
$\hat{k}_{y}\equiv-i\partial_{y}$ and $\hat{k}_{z}\equiv-i\partial_{z}$
are the momentum operators, and $\sigma_{x}$, $\sigma_{y}$ and $\sigma_{z}$
are the Pauli matrices. We assume a general form of 2D spin-orbit
coupling with $\lambda_{z}=\lambda\cos\theta$ and $\lambda_{x}=\lambda\sin\theta$,
where the case of $\theta=\pi/4$ corresponds to the Rashba spin-orbit
coupling. The system can be described by the model Hamiltonian, 
\begin{equation}
H=\int d{\bf x\left[{\cal H}_{\textrm{0}}\left({\bf x}\right)+{\cal {\cal H}_{\textrm{int}}}\left({\bf x}\right)\right]},
\end{equation}
where ${\cal H}_{0}(\mathbf{x})$ is the single-particle (density)
Hamiltonian 

\begin{equation}
{\cal H}_{0}=\left[\psi_{\uparrow}^{\dagger},\psi_{\downarrow}^{\dagger}\right]\left[\begin{array}{cc}
\hat{\xi}_{{\bf k}}+\lambda_{z}\hat{k}_{z}+h_{\textrm{in}} & \lambda_{x}\hat{k}_{x}-ih_{\textrm{out}}\\
\lambda_{x}\hat{k}_{x}+ih_{\textrm{out}} & \hat{\xi}_{{\bf k}}-\lambda_{z}\hat{k}_{z}-h_{\textrm{in}}
\end{array}\right]\left[\begin{array}{c}
\psi_{\uparrow}\\
\psi_{\downarrow}
\end{array}\right]
\end{equation}
and ${\cal {\cal H}_{\textrm{int}}}({\bf x})$ is the interaction
(density) Hamiltonian 
\begin{equation}
{\cal H}_{\textrm{int}}=U_{0}\psi_{\uparrow}^{\dagger}\left({\bf x}\right)\psi_{\downarrow}^{\dagger}\left({\bf x}\right)\psi_{\downarrow}\left({\bf x}\right)\psi_{\uparrow}\left({\bf x}\right)
\end{equation}
that describes the contact interaction between two atoms in different
spin states with an interaction strength $U_{0}$. In the above Hamiltonian,
$\psi_{\sigma}^{\dagger}\left({\bf x}\right)$ and $\psi_{\sigma}\left({\bf x}\right)$
are the creation and annihilation field operators of atoms in the
spin-state $\sigma=\uparrow,\downarrow$, respectively, and $\hat{\xi}_{\mathbf{k}}\equiv-\hbar^{2}\nabla^{2}/(2m)-\mu$
is the single-particle kinetic energy with atomic mass $m$, measured
with respect to the chemical potential $\mu$. The use of the contact
interatomic interaction necessarily leads to an ultraviolet divergence
at large momentum or high energy. To regularize it, the interaction
strength $U_{0}$ can be expressed in terms of an \textit{s}-wave
scattering length $a_{s}$, 
\begin{equation}
\frac{1}{U_{0}}=\frac{m}{4\pi\hbar^{2}a_{s}}-\frac{1}{V}\sum_{{\bf k}}\frac{m}{\hbar^{2}\mathbf{k}^{2}},
\end{equation}
where $V$ is the volume of the system. Experimentally, the scattering
length $a_{s}$ can be tuned precisely to arbitrary value, by sweeping
an external magnetic field across a Feshbach resonance \cite{Chin2010},
from the weakly interacting limit of Bardeen-Cooper-Schrieffer (BCS)
superfluids to the strong-coupling limit of Bose-Einstein condensates
(BEC) \cite{Giorgini2008}.

\subsection{Mean-field BdG theory}

It is now widely known that spin-orbit coupling together with in-plane
Zeeman field can induce the Fulde-Ferrell pairing, with an order parameter
having a finite center-of-mass momentum along the direction of in-plane
Zeeman field \cite{Zheng2013,Dong2013PRA,Wu2013,Liu2013a,Dong2013NJP,Zhou2013}.
This finite-momentum pairing - also referred to as the helical phase
- was first proposed in the study of noncentrosymmetric superconductors
\cite{Yip2014,Barzykin2002,Yip2002,Agterberg2007,Dimitrova2007,Michaeli2012}.
As we apply the in-plane Zeeman field along the $z$-direction, we
assume an order parameter 
\begin{equation}
\Delta\left(\mathbf{x}\right)=-U_{0}\left\langle \psi_{\downarrow}\left({\bf x}\right)\psi_{\uparrow}\left({\bf x}\right)\right\rangle =\Delta\exp\left(iqz\right)
\end{equation}
with a FF momentum $\mathbf{q}=q\mathbf{e}_{z}$, where $\mathbf{e}_{z}$
is the unit vector along the $z$-direction. Within the mean-field
theory, we may approximate the interaction Hamiltonian as
\begin{equation}
{\cal H}_{\textrm{int}}\simeq-\left[\Delta(\mathbf{x})\psi_{\uparrow}^{\dagger}(\mathbf{x})\psi_{\downarrow}^{\dagger}(\mathbf{x})+\textrm{H.c.}\right]-\frac{\Delta^{2}}{U_{0}}.
\end{equation}
For a Fermi superfluid, it is useful to introduce the following Nambu
spinor to collectively denote the field operators, 
\begin{equation}
\Phi(\mathbf{x})\equiv\left[\psi_{\uparrow}\left(\mathbf{x}\right),\psi_{\downarrow}\left(\mathbf{x}\right),\psi_{\uparrow}^{\dagger}\left(\mathbf{x}\right),\psi_{\downarrow}^{\dagger}\left(\mathbf{x}\right)\right]^{T},
\end{equation}
where the first two and last two field operators in $\Phi(\mathbf{x})$
could be interpreted as the annihilation operators for particles and
holes, respectively. The total mean-field Hamiltonian can then be
written in a compact form, 
\begin{equation}
H_{\textrm{mf}}=\sum_{\mathbf{k}}\hat{\xi}_{\mathbf{k}}-V\frac{\Delta^{2}}{U_{0}}+\frac{1}{2}\int d{\bf x}\Phi^{\dagger}(\mathbf{x})\mathcal{H}_{BdG}\Phi(\mathbf{x}),\label{eq:totalMFHami}
\end{equation}
where the factor of $1/2$ in the last term arises from the redundant
use of particle and hole operators in the Nambu spinor $\Phi(\mathbf{x})$.
Accordingly, a zero-point energy $\sum_{\mathbf{k}}\hat{\xi}_{\mathbf{k}}$
appears in the first term, whose divergence will be removed by regularizing
the bare interaction strength $U_{0}$ (see below). The Bogoliubov
Hamiltonian $\mathcal{H}_{BdG}$ takes the form,\begin{widetext}
\begin{equation}
{\cal H}_{BdG}\equiv\left[\begin{array}{cccc}
\hat{\xi}_{{\bf k}}+\lambda_{z}\hat{k}_{z}+h_{\textrm{in}} & \lambda_{x}\hat{k}_{x}-ih_{\textrm{out}} & 0 & -\Delta\left({\bf x}\right)\\
\lambda_{x}\hat{k}_{x}+ih_{\textrm{out}} & \hat{\xi}_{{\bf k}}-\lambda_{z}\hat{k}_{z}-h_{\textrm{in}} & \Delta\left({\bf x}\right) & 0\\
0 & \Delta^{*}\left({\bf x}\right) & -\hat{\xi}_{{\bf k}}+\lambda_{z}\hat{k}_{z}-h_{\textrm{in}} & \lambda_{x}\hat{k}_{x}-ih_{\textrm{out}}\\
-\Delta^{*}\left({\bf x}\right) & 0 & \lambda_{x}\hat{k}_{x}+ih_{\textrm{out}} & -\hat{\xi}_{{\bf k}}-\lambda_{z}\hat{k}_{z}+h_{\textrm{in}}
\end{array}\right].\label{eq:BdGHami}
\end{equation}
 \end{widetext} Note that, the spin-orbit coupling term in the hole
sector of Eq. (\ref{eq:BdGHami}) does not change sign as spin-orbit
coupling respects the time-reversal symmetry. The mean-field Hamiltonian
Eq. (\ref{eq:totalMFHami}) can be solved by taking the standard Bogoliugov
transformation. In our case, it is more convenient to directly diagonalizing
the Bogoliubov Hamiltonian 
\begin{equation}
\mathcal{H}_{BdG}\Phi_{\mathbf{k\eta}}(\mathbf{x})=E_{\mathbf{\eta}}\left(\mathbf{k}\right)\Phi_{\mathbf{k\eta}}(\mathbf{x})\label{eq:BogoliubovEQ}
\end{equation}
 with the plane-wave quasiparticle wave-function
\begin{equation}
\Phi_{{\bf k}\eta}\left({\bf x}\right)\equiv\frac{e^{i{\bf k\cdot x}}}{\sqrt{V}}\left[\begin{array}{c}
u_{{\bf k\eta}\uparrow}e^{+iqz/2}\\
u_{{\bf k\eta}\downarrow}e^{+iqz/2}\\
v_{{\bf k}\eta\uparrow}e^{-iqz/2}\\
v_{{\bf k}\eta\downarrow}e^{-iqz/2}
\end{array}\right]\label{eq:BogoliubovWF}
\end{equation}
and quasiparticle energy $E_{\eta}(\mathbf{k})$. As the Bogoliubov
Hamiltonian now becomes a 4 by 4 matrix, the four eigenvalues and
eigenstates can be collectively specified as $\eta=\{\alpha,\nu\}$,
where the index $\alpha\in(1,2)$ indicates the upper ($1$) or lower
($2$) band split by the spin-orbit coupling and Zeeman fields, and
$\nu\in(+,-)$ stands for the particle ($+$) or hole ($-$) branch.
The ansatz of the wave-fucntion Eq. (\ref{eq:BogoliubovWF}) is inspired
by the use of the Fulde-Ferrell order parameter $\Delta\exp(iqz)$
in the Bogoliubov Hamiltonian and is also consistent with the fact
that the hole wave-function is a time-reversal partner of the particle
wave-fucntion. By substituting Eq. (\ref{eq:BogoliubovWF}) into the
Bogoliubov equation (\ref{eq:BogoliubovEQ}), we obtain a $4\times4$
Bogoliubov matrix for $[u_{\mathbf{k}\eta\uparrow},u_{\mathbf{k\eta}\downarrow},v_{\mathbf{k\eta}\uparrow},v_{\mathbf{k\eta}\downarrow}]$$^{T}$,
in which the operators $\hat{\xi}_{{\bf k}}$ is replaced with a $c$-number
\begin{equation}
\xi_{\mathbf{k}+\mathbf{q}/2}=\frac{\hbar^{2}}{2m}\left[k_{x}^{2}+k_{y}^{2}+\left(k_{z}+\frac{q}{2}\right)^{2}\right]-\mu
\end{equation}
and $\hat{k}_{z}$ replaced with $k_{z}+q/2$ for the particle sector
and, similarly 
\begin{equation}
\hat{\xi}_{{\bf k}}\rightarrow\xi_{\mathbf{k}-\mathbf{q}/2}=\frac{\hbar^{2}}{2m}\left[k_{x}^{2}+k_{y}^{2}+\left(k_{z}-\frac{q}{2}\right)^{2}\right]-\mu
\end{equation}
and $\hat{k}_{z}\rightarrow k_{z}-q/2$ for the hole sector (see,
for example, Eq. (\ref{eq:BdGMatrix}) below with $-i\partial_{x}\rightarrow k_{x}$).
Once the quasiparticle energy $E_{\eta}(\mathbf{k})$ is known by
diagonalizing the (complex) Bogoliubov matrix, we obtain the mean-field
thermodynamic potential $\Omega_{\textrm{mf}}$ at a finite temperature
$T$: 
\begin{eqnarray}
\frac{\Omega_{\textrm{mf}}}{V} & = & \frac{1}{2V}\left[\sum_{\mathbf{k}}\left(\xi_{\mathbf{k}+\mathbf{q}/2}+\xi_{\mathbf{k}-\mathbf{q}/2}\right)-\sum_{\mathbf{k\eta}}E_{\eta}\left(\mathbf{k}\right)\right]-\frac{\Delta^{2}}{U_{0}}\nonumber \\
 &  & -\frac{k_{B}T}{V}\sum_{\mathbf{k\eta}}\ln\left[1+e^{-E_{\eta}\left(\mathbf{k}\right)/k_{B}T}\right],\label{eq:Omf}
\end{eqnarray}
where the zero-point energy $(1/2)\sum_{\mathbf{k\eta}}E_{\eta}(\mathbf{k})$
in the square bracket is again due to the redundant use of particle
and hole operators, and we have rewritten the zero-point energy $\sum_{\mathbf{k}}\hat{\xi}_{\mathbf{k}}$
to the form of $\sum_{\mathbf{k}}(\xi_{\mathbf{k}+\mathbf{q}/2}+\xi_{\mathbf{k}-\mathbf{q}/2})/2$,
in order to cancel the leading divergence in $(1/2)\sum_{\mathbf{k\eta}}E_{\eta}(\mathbf{k})$.
The last term in the thermodynamic potential accounts for the thermal
excitations of Bogoliubov quasiparticles, which are non-interacting
within the mean-field approximation. It worth noting that, the summation
over the quasiparticle energy in $\sum_{\mathbf{k\eta}}$ must be
restricted to $E_{\eta}(\mathbf{k})\geq0$, because of the inherent
particle-hole symmetry of the Bogoliubov Hamiltonian.

\subsection{Chiral surface states and Majorana fermions}

In the topological phase, the superfluid can support gapless surface
states. To demonstrate this non-trivial consequence of topological
order, we consider adding a hard-wall confinement in the $x$-direction
and calculate the resulting energy spectrum $E_{n}(k_{z},k_{y})$.
The zero-energy Majorana fermion modes are expected to appear at $k_{z}=0$
and at the two boundaries $x=0$ and $x=L$ \cite{Wilczek2009,Hasan2010,Qi2011},
where $L\gg k_{F}^{-1}$ is the length of the confinement along the
$x$-direction.

The use of a hard-wall potential means that the wave-function of quasiparticles
along the $x$-axis is no longer a plane wave. By assuming that the
order parameter is approximately not affected by the confinement for
sufficiently large $L$, we take the ansatz,
\begin{equation}
\Phi_{k_{y}k_{z}}\left({\bf x}\right)\equiv\frac{e^{i\left(k_{y}y+k_{z}z\right)}}{\sqrt{V}}\left[\begin{array}{c}
u_{k_{y}k_{z}\uparrow}\left(x\right)e^{+iqz/2}\\
u_{k_{y}k_{z}\downarrow}\left(x\right)e^{+iqz/2}\\
v_{k_{y}k_{z}\uparrow}\left(x\right)e^{-iqz/2}\\
v_{k_{y}k_{z}\downarrow}\left(x\right)e^{-iqz/2}
\end{array}\right].\label{eq:ChrialSurfaceStatesWF}
\end{equation}
By substituting Eq. (\ref{eq:ChrialSurfaceStatesWF}) into Eq. (\ref{eq:BdGHami}),
the resulting BdG Hamiltonian for a given set of ($k_{z},k_{y}$)
has the form,\begin{widetext} 
\begin{equation}
{\cal H}_{BdG}=\left[\begin{array}{cccc}
-\frac{\hbar^{2}}{2m}\partial_{x}^{2}+\tilde{H}_{11}^{(p)} & i\left(-\lambda_{x}\partial_{x}-h_{\textrm{out}}\right) & 0 & -\Delta\\
i\left(-\lambda_{x}\partial_{x}+h_{\textrm{out}}\right) & -\frac{\hbar^{2}}{2m}\partial_{x}^{2}+\tilde{H}_{22}^{(p)} & \Delta & 0\\
0 & \Delta & +\frac{\hbar^{2}}{2m}\partial_{x}^{2}+\tilde{H}_{11}^{(h)} & i\left(-\lambda_{x}\partial_{x}-h_{\textrm{out}}\right)\\
-\Delta & 0 & i\left(-\lambda_{x}\partial_{x}+h_{\textrm{out}}\right) & +\frac{\hbar^{2}}{2m}\partial_{x}^{2}+\tilde{H}_{22}^{(h)}
\end{array}\right],\label{eq:BdGMatrix}
\end{equation}
where 
\begin{eqnarray}
\tilde{H}_{11}^{(p)} & = & +\left[\frac{\hbar^{2}}{2m}\left(k_{y}^{2}+k_{z}^{2}+\frac{q^{2}}{4}\right)-\mu\right]+\frac{\hbar^{2}}{2m}qk_{z}+\lambda_{z}\left(k_{z}+\frac{q}{2}\right)+h_{\textrm{in}},\\
\tilde{H}_{22}^{(p)} & = & +\left[\frac{\hbar^{2}}{2m}\left(k_{y}^{2}+k_{z}^{2}+\frac{q^{2}}{4}\right)-\mu\right]+\frac{\hbar^{2}}{2m}qk_{z}-\lambda_{z}\left(k_{z}+\frac{q}{2}\right)-h_{\textrm{in}},\\
\tilde{H}_{11}^{(h)} & = & -\left[\frac{\hbar^{2}}{2m}\left(k_{y}^{2}+k_{z}^{2}+\frac{q^{2}}{4}\right)-\mu\right]+\frac{\hbar^{2}}{2m}qk_{z}+\lambda_{z}\left(k_{z}-\frac{q}{2}\right)-h_{\textrm{in}},\\
\tilde{H}_{22}^{(h)} & = & -\left[\frac{\hbar^{2}}{2m}\left(k_{y}^{2}+k_{z}^{2}+\frac{q^{2}}{4}\right)-\mu\right]+\frac{\hbar^{2}}{2m}qk_{z}-\lambda_{z}\left(k_{z}-\frac{q}{2}\right)+h_{\textrm{in}}.
\end{eqnarray}
\end{widetext} To diagonalize the BdG Hamiltonian Eq. (\ref{eq:BdGMatrix}),
we expand all the quasiparticle wavefunctions in terms of the single-particle
eigenstates of the hard-wall potential, which is given by 
\begin{equation}
\psi_{l}\left(x\right)=\sqrt{\frac{2}{L}}\sin\left(\frac{l\pi x}{L}\right),
\end{equation}
with an integer $l=1,2,3,\cdots$. For example, for $u_{k_{y}k_{z}\uparrow}\left(x\right)$
we have 
\begin{equation}
u_{k_{y}k_{z}\uparrow}\left(x\right)=\sum_{l=1}^{N_{\max}}u_{k_{y}k_{z}\uparrow}^{\left(l\right)}\psi_{l}\left(x\right),
\end{equation}
where $N_{\max}\gg1$ is a high energy cut-off. This converts the
BdG Hamiltonian Eq. (\ref{eq:BdGMatrix}) into a $4N_{\max}$ by $4N_{\max}$
symmetric matrix, with the matrix elements, 
\begin{equation}
\left[-\partial_{x}^{2}\right]_{lm}=\left(\frac{\pi l}{L}\right)^{2}\delta_{lm}
\end{equation}
and
\begin{equation}
\left[\partial_{x}\right]_{lm}=\frac{m}{L}\left[\frac{1-\cos\pi\left(l+m\right)}{\left(l+m\right)}+\frac{1-\cos\pi\left(l-m\right)}{\left(l-m\right)}\right].
\end{equation}
The diagonalization leads directly to the energies $E_{n}(k_{y},k_{z})$
and wavefunctions of the chiral surface states.

\section{Results and discussions}

For a uniform Fermi gas, we use the natural units, $2m=\hbar=k_{B}=k_{F}=E_{F}=1$.
This means that we take the Fermi wavevector $k_{F}$ and Fermi energy
$E_{F}$ as the units for wavevector and energy. Thus, the spin-orbit
coupling will be parameterized by the dimensionless parameter $\tilde{\lambda}_{z,x}=\lambda_{z,x}k_{F}/E_{F}$.
It is also understood that for the Zeeman fields we use $\tilde{h}_{\textrm{in}}=h_{\textrm{in}}/E_{F}$
and $\tilde{h}_{\textrm{out}}=h_{\textrm{out}}/E_{F}$. To find possible
mean-field phases, for a given set of parameters (including spin-orbit
coupling $\lambda k_{F}/E_{F}$, Zeeman field $h/E_{F}$, interaction
parameter $1/(k_{F}a_{s})$, and temperature $T/T_{F}$) and total
density $N/V=k_{F}^{3}/(3\pi^{2})=1/(3\pi^{2})$, we need to minimize
the mean-field thermodynamic potential against $\Delta$ and $q$,
i.e., 
\begin{eqnarray}
\frac{\partial\Omega_{\textrm{mf}}}{\partial\Delta} & = & 0,\\
\frac{\partial\Omega_{\textrm{mf}}}{\partial q} & = & 0.
\end{eqnarray}
Moreover, we have the number equation, 
\begin{equation}
-\frac{1}{V}\frac{\partial\Omega_{\textrm{mf}}}{\partial\mu}=\frac{N}{V}=\frac{1}{3\pi^{2}}.
\end{equation}
These equations are solved self-consistently by using Newton's gradient
approach. The various derivatives can be approximated by numerical
differences. The approach is very efficient, provided a good guess
of the initial parameters, which may be obtained by plotting the contour
plot of the mean-field thermodynamic potential. Throughout the paper,
we focus on the BCS side with a dimensionless interaction parameter
$1/(k_{F}a_{s})=-0.5$ and use mean-field theory to solve the model
Hamiltonian at both zero and finite temperatures. The spin-orbit coupling
strength is always fixed to $\lambda=E_{F}/k_{F}$. In most cases,
we consider a Rashba spin-orbit coupling. The deviation from the Rashba
case will be investigated at the end of the section.

\subsection{Zero temperature phase diagram}

\subsubsection{Out-of-plane Zeeman field}

\begin{figure}
\begin{centering}
\includegraphics[clip,width=0.48\textwidth]{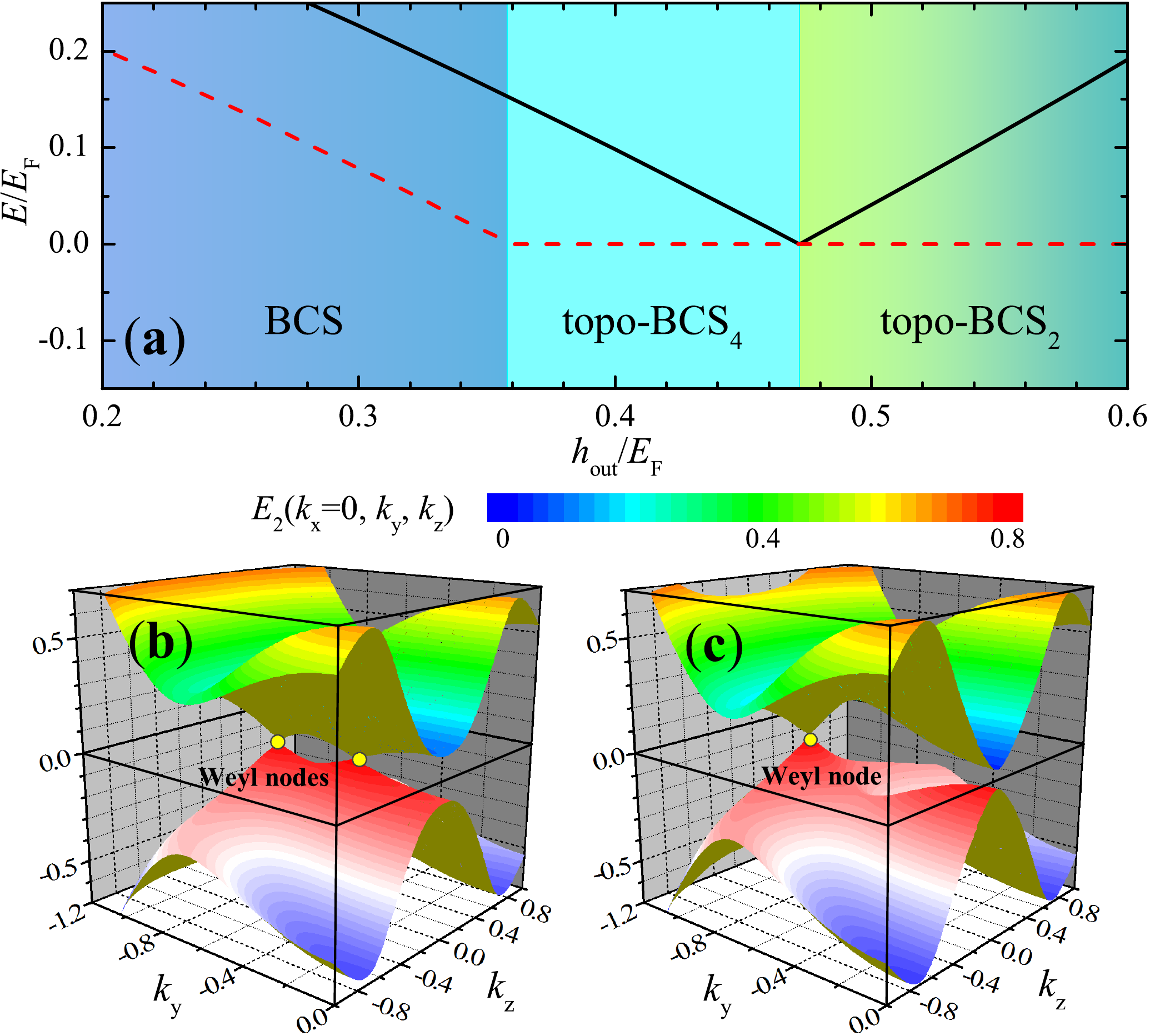} 
\par\end{centering}

\protect\caption{(Color online). (a) Evolution of the superfluid phase with increasing
the out-of-plane Zeeman field, at zero in-plane Zeeman field $h_{\textrm{in}}=0$.
The system evolves from a standard BCS superfluid to a topological
non-trivially superfluid with either four (topo-BCS$_{4}$) or two
Weyl nodes (topo-BCS$_{2}$). The solid and dashed lines show the
energy at $\mathbf{k}=0$ {[}$E_{2+}(\mathbf{k}=0)${]} and the minimum
energy {[}$\min E_{2+}(\mathbf{k})${]} of the lower particle branch.
(b) and (c) show the characteristic quasiparticle excitation spectrum
$E_{2\pm}(k_{x}=0,k_{y},k_{z})$ in the topo-BCS$_{4}$ ($h_{\textrm{out}}=0.4E_{F}$)
and topo-BCS$_{2}$ phases ($h_{\textrm{out}}=0.5E_{F}$), respectively.
We note that the spectrum is symmetric with respect to $k_{y}$ and
we plot only the left part with $k_{y}\leq0$.}

\label{fig2} 
\end{figure}

\begin{figure}
\begin{centering}
\includegraphics[clip,width=0.48\textwidth]{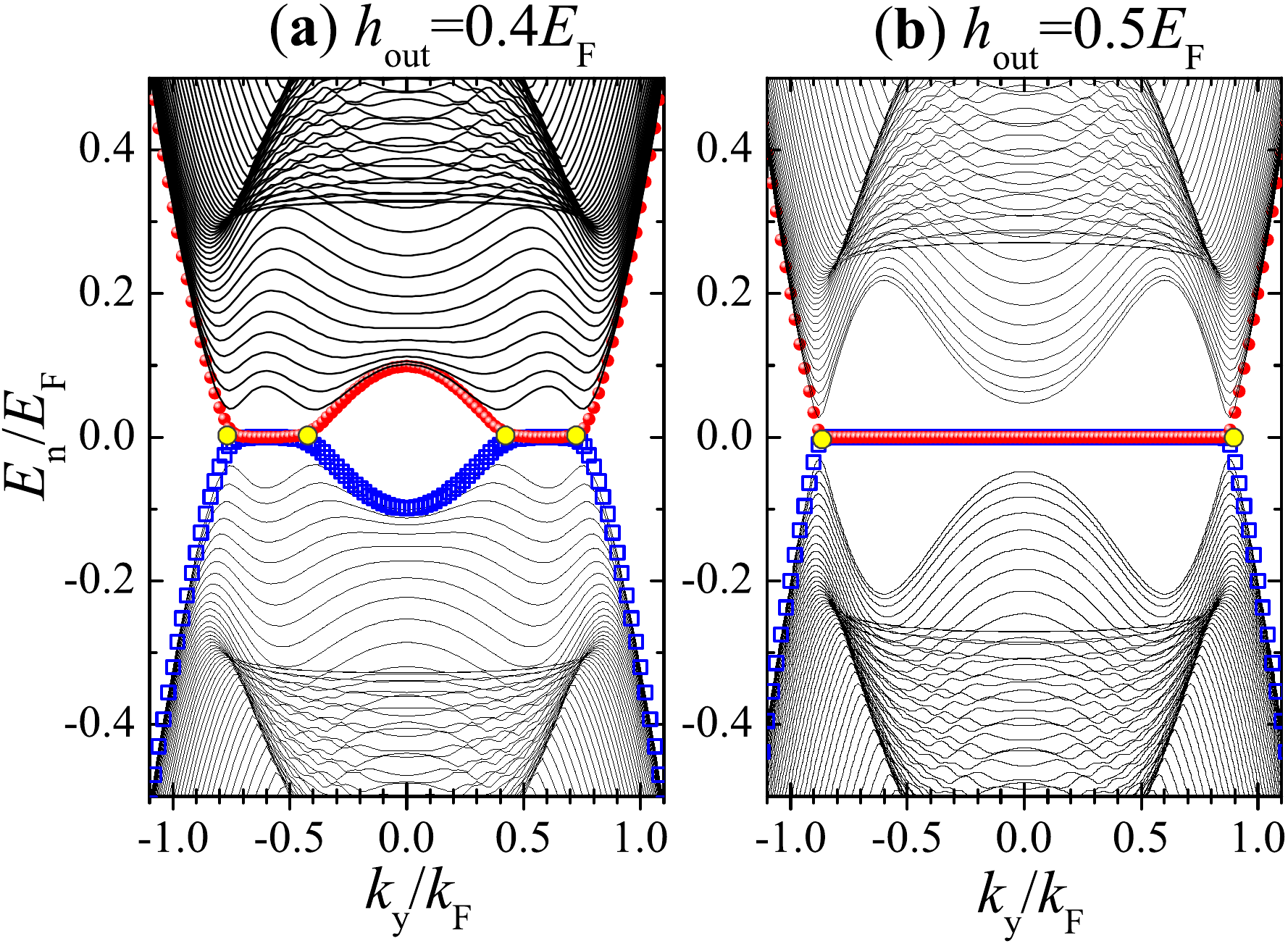} 
\par\end{centering}

\protect\caption{(Color online). Chiral surface modes - highlighted by blue empty squares
and red solid circles - with a hard-wall confinement along the \textit{x}-direction
at $k_{z}=0$, in the topo-BCS$_{4}$ (a) or topo-BCS$_{2}$ phases
(b). The system may be viewed as a stack of many 2D topological superfluids
along the \textit{y}-direction. Accordingly, the surface modes of
the system may be understood as a collection of the edge modes of
these 2D systems.}

\label{fig3} 
\end{figure}

Let us first consider the simplest case with an out-of-plane Zeeman
field only, a situation that was previously addressed by Gong and
co-workers \cite{Gong2011}. The zero-temperature phase diagram is
reported in Fig. \ref{fig2}(a), together with the energy spectra
of two topologically non-trivial phases at $h_{\textrm{out}}=0.4E_{F}$
and $h_{\textrm{out}}=0.5E_{F}$ in Figs. \ref{fig2}(b) and \ref{fig2}(c),
respectively. The chiral surface states in the presence of a hard-wall
confinement at the two boundaries along the $x$-direction are shown
in Fig. \ref{fig3}.

In the phase diagram, the properties of any possible 3D phase can
be understood by \emph{dimensional reduction} in momentum space to
a set of 2D phases \cite{Hasan2010,Sau2012}. The basic idea of dimensional
reduction is to treat the 3D Hamiltonian, for example, $\mathcal{H}_{BdG}(\hat{k}_{x,}\hat{k}_{y},\hat{k}_{z})$
of the 3D bulk system, Eq. (\ref{eq:BdGHami}) or Eq. (\ref{eq:BdGMatrix})
- which is translationally invariant along the $y$-direction (and
therefore the momentum operator $\hat{k}_{y}$ can simply be replaced
with a $c$-number $k_{y}$) - as a set of 2D systems, i.e., 
\begin{equation}
\mathcal{H}_{BdG}(\hat{k}_{x,}\hat{k}_{y},\hat{k}_{z})=\sum_{k_{y}}\mathcal{H}_{k_{y}}^{(2D)}\left(\hat{k}_{x,},\hat{k}_{z}\right),\label{eq:dimReduction}
\end{equation}
each of which is described by an effective 2D Hamiltonian $\mathcal{H}_{k_{y}}^{(2D)}(\hat{k}_{x,},\hat{k}_{z})$
parameterized by $k_{y}$. In the context of topological insulators
and superfluids \cite{Hasan2010}, this idea was successfully used
to understand the classification of 3D topological states. A \emph{weak}
3D topological phase can be defined, as a stack of 2D topological
phases, which are \emph{dispersionless} along the $y$-direction.
Actually, this is exactly the case happened in our 3D system. We claim
the existence of a 3D topological phase if there is an effective 2D
Hamiltonian with suitable $k_{y}$ that can host a topological phase.
The chiral surfaces states of the 3D topological phase may simply
be understood as the edge states of many 2D systems accumulated along
the $y$-axis and characterized by a good quantum number $k_{y}$
(see Fig. \ref{fig3}). The dispersion of these surface states in
other directions can be obtained from the dispersion of the edge states
\cite{Sau2012}. It is worth noting that this scenario of dimensional
reduction works for arbitrary in-plane and out-of-plane Zeeman fields.
Numerically, it is convenient to identify a 3D topological phase by
looking at whether there are chiral surface states or not.

In the presence of an out-of-plane Zeeman field $h_{\textrm{out}}$
only, it is known that the 2D Hamiltonian $\mathcal{H}_{k_{y}}^{(2D)}(\hat{k}_{x,},\hat{k}_{z})$
supports a topological phase when the field strength $h_{\textrm{out}}$
is above a threshold \cite{Hasan2010,Zhu2011,Liu2012a}. The topological
phase transition is accompanied by the closing and re-opening of the
bulk quasiparticle excitation gap at $k_{x}=0$ and $k_{z}=0$, which
from Eq. (\ref{eq:BdGHami}) is given by \cite{Gong2011},
\begin{equation}
E_{g}=2\left|h_{\textrm{out}}-\sqrt{\left[\mu\left(k_{y}\right)\right]^{2}+\Delta^{2}}\right|,
\end{equation}
where $\mu(k_{y})\equiv\mu-\hbar^{2}k_{y}^{2}/(2m)$ is a ``local''
chemical potential for the slice with $k_{y}$ in momentum space,
and the chemical potential $\mu$ and pairing gap $\Delta$ themselves
are functions of the out-of-plane Zeeman field $h_{\textrm{out}}$.
Thus, the transition occurs at the threshold 
\begin{equation}
h_{\textrm{out}}^{(c)}=\sqrt{\left[\mu\left(k_{y}\right)\right]^{2}+\Delta^{2}},
\end{equation}
whose value depends critically on $k_{y}$. Due to a positive chemical
potential $\mu$ on the BCS side, it is easy to see that the topological
phase transition first occurs at $\pm k_{y}^{(c)}\neq0$. 

By increasing the out-of-plane Zeeman field, the slices with $k_{y}\subset[-k_{2}^{(W)},-k_{1}^{(W)}]\cup[k_{1}^{(W)},k_{2}^{(W)}]$
become topologically non-trivial (see, for example, Fig. \ref{fig3}(a)).
The momenta $\pm k_{1}^{(W)}$ and $\pm k_{2}^{(W)}$, satisfying
$k_{1}^{(W)}<k_{y}^{(c)}<k_{2}^{(W)}$, are four boundary points at
which the excitation gap must close. These are the so-called Weyl
nodes, around each of which the effective Hamiltonian describes quasiparticles
with a linear dispersion that resembles a 3D Dirac cone (see Fig.
\ref{fig2}(b) for the two Weyl nodes at $k_{y}<0$). These Dirac
cone spectra are very robust against any perturbation of the bulk
Hamiltonian $\mathcal{H}_{BdG}$ and are topologically protected by
a topological invariant \cite{Gong2011,Sau2012}. They are exactly
an ultracold-atom analog of Weyl fermions of particle physics. In
this respect, we may refer to the topological phase as a Weyl superfluid.
Due to the existence of Weyl nodes, we see immediately that the global
energy gap in the topological phase must vanish. By further increasing
the out-of-plane Zeeman field, the lower bound momentum $k_{1}^{(W)}$
may become zero and therefore the two inner Weyl nodes (with \emph{opposite}
topological invariant) annihilate with each other. This happens when
the local energy gap at $\mathbf{k}=0$ closes. Afterwards, the system
supports two Weyl nodes.

From the above analysis, it is clear that the appearance of a topological
phase with four and two Weyl nodes can be determined by the closing
of the global energy gap and of the local energy gap at $\mathbf{k}=0$,
respectively. This is demonstrated in Fig. \ref{fig2}(a), where we
plot the minimum energy {[}$\min E_{2+}(\mathbf{k})${]} and the local
energy at $\mathbf{k}=0$ {[}$E_{2+}(\mathbf{k}=0)${]} of the lower
particle branch by using the dashed and solid lines, respectively.
Here we use the minimum energy, instead of a global energy gap $E_{g}=2\left|\min E_{2+}(\mathbf{k})\right|$
that is the energy difference between the minimum energy of the particle
branch and the maximum of the hole branch due to the particle-hole
symmetry $E_{2+}(\mathbf{k})=-E_{2-}(-\mathbf{k})$. As we shall see
below, the minimum energy is more useful to characterize the emergence
of gapless phase \cite{Hu2014}. With increasing out-of-plane Zeeman
field, the minimum energy develops a zero-energy \emph{plateau} as
the system enters the topological phase, due to the occasional band
touching of the particle and hole branches at Weyl nodes. In contrast,
the local energy at $\mathbf{k}=0$ will close and then reopen at
the critical field where the number of Weyl nodes decreases from four
to two. 

In the topological phase, the formation of isolated Weyl nodes can
be clearly identified in the energy spectrum in momentum space, as
shown in Figs. \ref{fig2}(b) and \ref{fig2}(c). It is interesting
that these Weyl nodes are connected continuously by zero-energy Majorana
surface states once we impose the hard-wall confinement at the two
boundaries along the $x$-direction (see Fig. \ref{fig3}). The existence
of the Majorana modes is easy to understand from the dimensional reduction
mentioned earlier: the surface states with a given $k_{y}$ and $k_{z}$
of the 3D Hamiltonian $\mathcal{H}_{BdG}$ is a sum of the edge state
with momentum $k_{z}$ of the 2D Hamiltonian in Eq. (\ref{eq:dimReduction})
with $k_{y}$ as a parameter value. For $k_{z}=0$, all the edge states
have zero-energy at any value of $k_{y}$ in the range where the 2D
slice is topologically nontrivial and, therefore, they form a Majorana
flat band \cite{Sau2012,Wong2013}.

\begin{figure}[t]
\begin{centering}
\includegraphics[clip,width=0.48\textwidth]{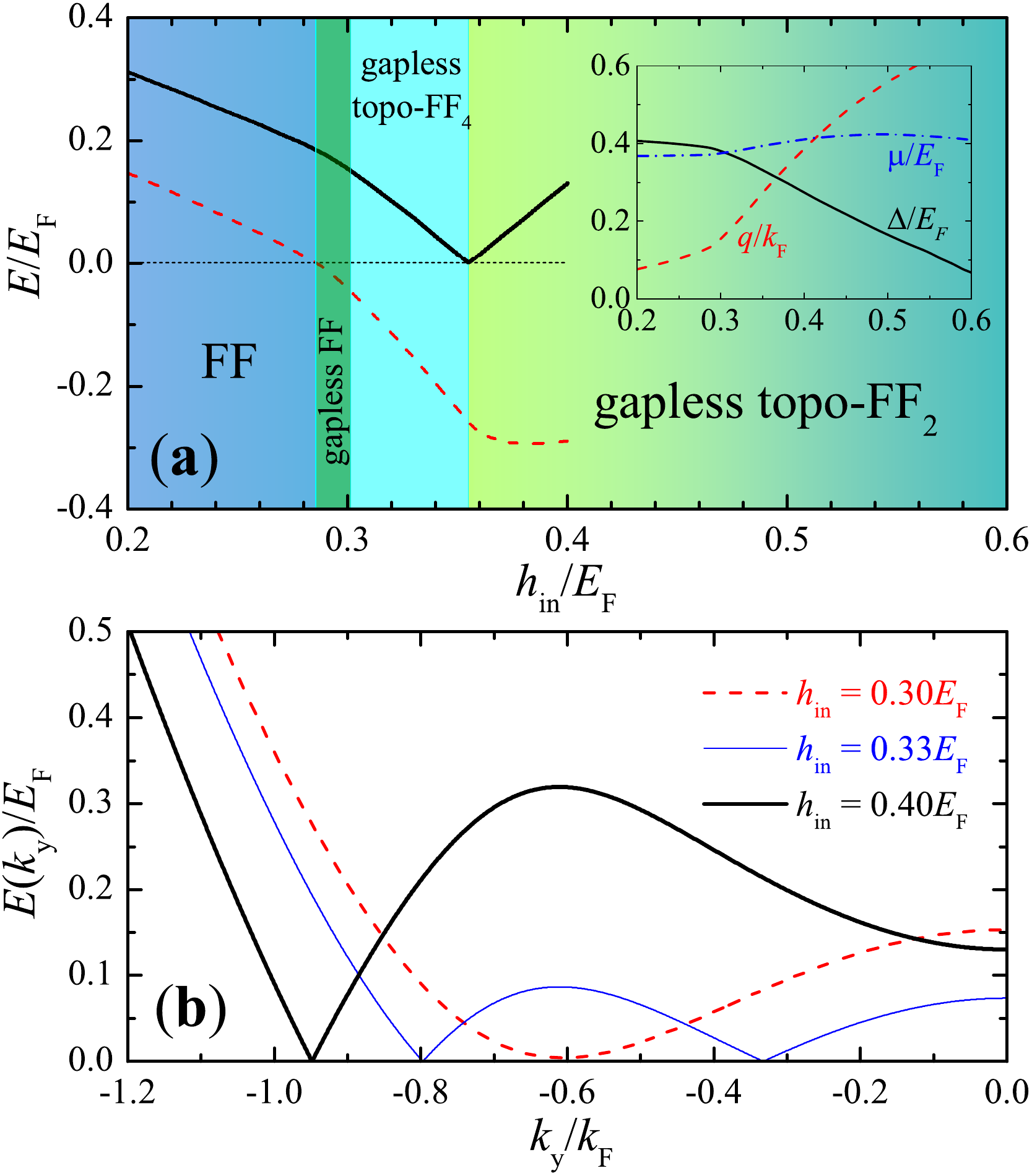} 
\par\end{centering}

\protect\caption{(Color online). (a) Evolution of the superfluid phase with increasing
the in-plane Zeeman field, at zero out-of-plane Zeeman field $h_{\textrm{out}}=0$.
Driven by the in-plane field only, the system evolves from a gapped
FF superfluid to a gapless FF state, and then to a topologically non-trivial
FF superfluid with either four (gapless topo-FF$_{4}$) or two Weyl
nodes/loops (gapless topo-FF$_{2}$). The solid and dashed lines show
the energy at $\mathbf{k}=0$ {[}$E_{2+}(\mathbf{k}=0)${]} and the
minimum energy {[}$\min E_{2+}(\mathbf{k})${]} of the lower particle
branch. The inset shows the chemical potential, pairing gap and FF
pairing momentum as a function of the in-plane Zeeman field. (b) The
energy of the lower particle branch along the $k_{y}$-direction,
$E_{2+}(k_{x}=0,k_{y},k_{z}=0)$, at $h_{\textrm{in}}/E_{F}=0.30$
(gapless FF, dashed line), $0.33$ (gapless topo-FF$_{4}$, thin solid
line) and $0.40$ (gapless topo-FF$_{2}$, thick solid line). The
minimum energy touches zero at the positions of Weyl nodes.}

\label{fig4} 
\end{figure}

To close this subsection, we note that, with a small in-plane Zeeman
field, the phase diagram is essentially unchanged. The only difference
is that Cooper pairs now acquire a finite center-of-mass momentum.
The resulting topologically nontrivial phases are better referred
to as topo-FF$_{4}$ or topo-FF$_{2}$, according to the number of
Weyl nodes. The linear quasiparticle spectrum around a Weyl node becomes
anisotropic \cite{Xu2014}. In the presence of a large in-plane Zeeman
field, however, the phase diagram would change qualitatively, as we
now turn to discuss in great detail.

\subsubsection{In-plane Zeeman field}

Figure \ref{fig4}(a) reports the phase diagram in the presence of
an in-plane Zeeman field only. Compared with Fig. \ref{fig2}, with
increasing the Zeeman field strength, the local energy at $\mathbf{k}=0$
(solid line) shows similar behavior. On the contrary, the minimum
energy $\min E_{2+}(\mathbf{k})$ (dashed line) exhibits very different
field dependence. It becomes negative at $h_{\textrm{in}}>h_{\textrm{in}}^{(gapless)}\simeq0.284E_{F}$
and indicates clearly that the spin-orbit coupled Fermi gas enters
a gapless phase, due to the existence of a significant Fulde-Ferrell
pairing momentum at the order of Fermi momentum (i.e., $q\sim k_{F}$,
see the inset). It is natural to ask: would the gapless system become
topologically non-trivial at a larger in-plane Zeeman field $h_{\textrm{in}}^{(c)}>h_{\textrm{in}}^{(gapless)}$
?

\begin{figure}
\begin{centering}
\includegraphics[clip,width=0.48\textwidth]{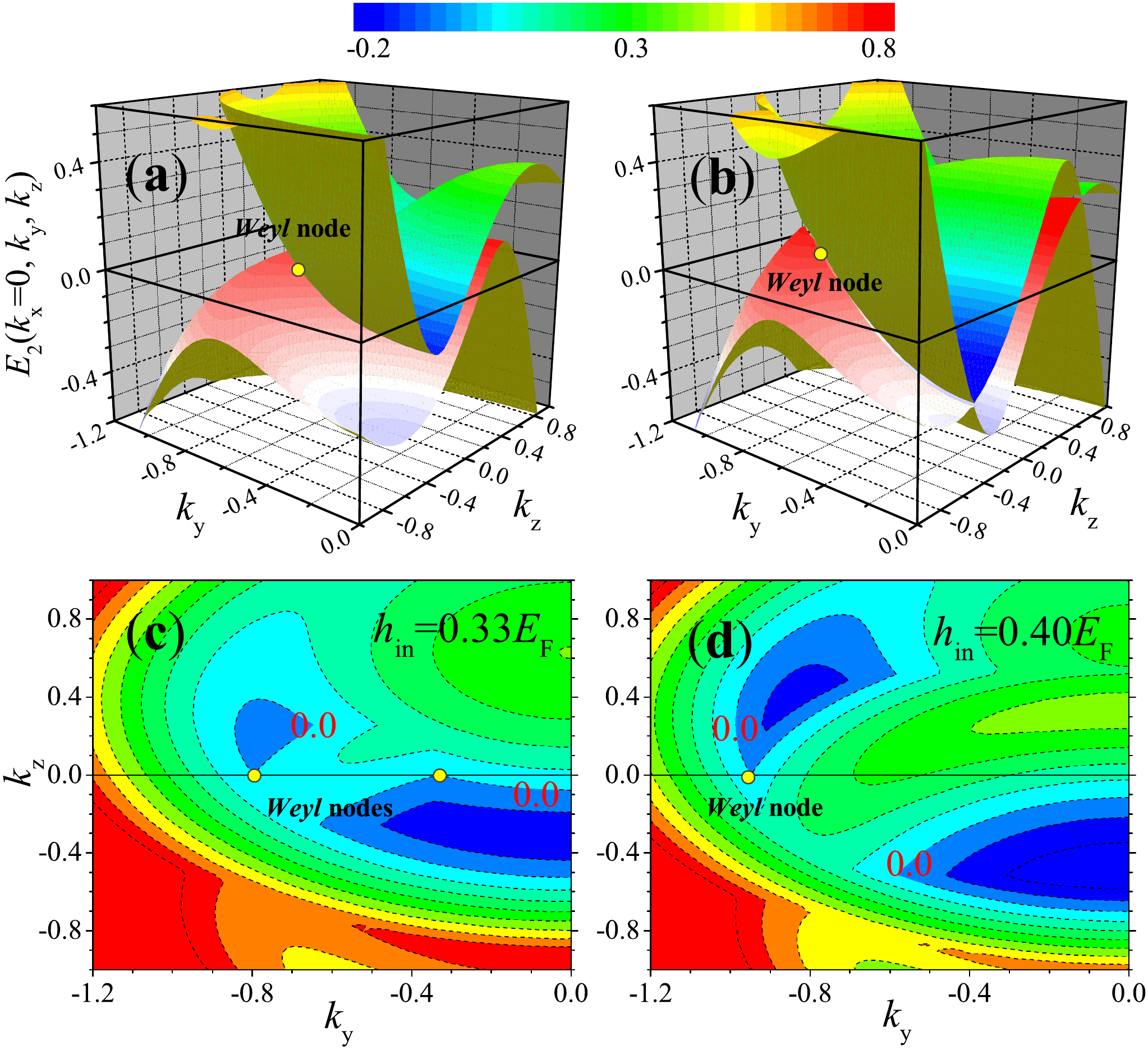} 
\par\end{centering}

\protect\caption{(Color online). The characteristic quasiparticle excitation spectrum
$E_{2\pm}(k_{x}=0,k_{y},k_{z})$ in the gapless topo-FF$_{4}$ phase
with $h_{\textrm{in}}=0.33E_{F}$ (a) and in the gapless topo-FF$_{2}$
phase with $h_{\textrm{in}}=0.40E_{F}$ (b). The corresponding contour
plots of $E_{2+}(k_{x}=0,k_{y},k_{z})$ are shown in (c) and (d),
respectively. The Weyl nodes are highlighted by yellow solid circles.
We note that the out-of-plane Zeeman field $h_{\textrm{out}}=0$.
The spectrum is symmetric with respect to $k_{y}$, so we plot only
the left part with $k_{y}\leq0$.}

\label{fig5} 
\end{figure}

\begin{figure*}
\begin{centering}
\includegraphics[clip,width=0.7\textwidth]{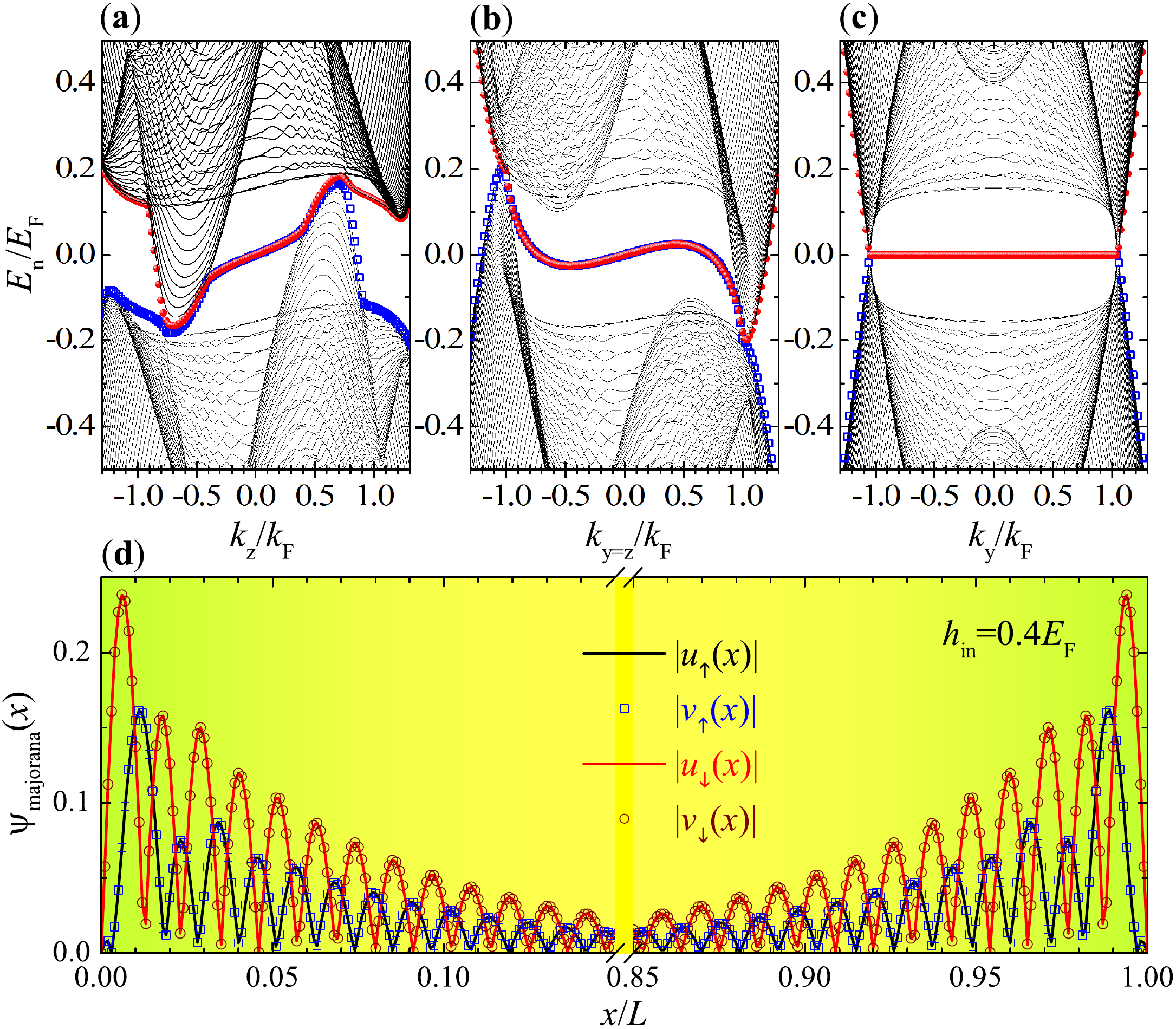} 
\par\end{centering}

\protect\caption{(Color online). Chiral surface modes with a hard-wall confinement
along the \textit{x}-direction in the gapless topo-FF$_{2}$ phase
($h_{\textrm{in}}=0.4E_{F}$), at $k_{y}=0$ (a), along the $k_{y}=k_{z}$
line (b), or at $k_{z}=0$ (c). The wave-functions of the Majorana
mode at $k_{y}=k_{z}=0$ is shown in (d), which are well localized
at the two boundaries $x=0$ and $x=L=200k_{F}^{-1}$.}

\label{fig6} 
\end{figure*}
Our answer is yes, after a careful check on the structure of the energy
spectrum (Fig. \ref{fig5}) as well as the appearance of the Majorana
surface states (Fig. \ref{fig6}). The energy spectrum can have Weyl
nodes as the in-plane Zeeman field $h_{\textrm{in}}>h_{\textrm{in}}^{(c1)}\simeq0.301E_{F}$.
This is particularly clear when we plot the spectrum at $k_{x}=k_{z}=0$
as a function of the parameterized momentum $k_{y}$ \cite{footnote1},
as shown in Fig. \ref{fig4}(b). Resembling the earlier case with
an out-of-plane Zeeman field, four Weyl nodes develop once $h_{\textrm{in}}>h_{\textrm{in}}^{(c1)}$
in the gapless topo-FF$_{4}$ phase. In addition, by further increasing
the in-plane Zeeman field above a threshold $h_{\textrm{in}}^{(c2)}\simeq0.355E_{F}$,
exactly at which the local energy gap at $\mathbf{k}=0$ closes, the
number of nodes decreases to two in the gapless topo-FF$_{2}$ phase.
There is, however, a notable difference when we examine the full energy
spectrum in Fig. \ref{fig5}. In the gapless phase, the nodal points
will enclose and form two surfaces, one in the particle branch and
another in the hole branch \cite{Dong2013NJP}. Along the cut $k_{x}=0$,
as shown in Figs. \ref{fig5}(c) and \ref{fig5}(d), each Weyl node
will then extend to form a closed loop (i.e., Weyl loop). Nevertheless,
the occasional band touching of the particle and hole branches still
occurs at a single point on the $k_{z}=0$ axis and therefore the
topological nature of the spectrum remains unchanged.

In Figs. \ref{fig6}(a), \ref{fig6}(b) and \ref{fig6}(c), we present
the energy dispersion of the surface states in the gapless topo-FF$_{2}$
phase. At $k_{z}=0$ (Fig. \ref{fig6}(c)), we observe the same Majorana
flat band as in the case of out-of-plane Zeeman field (see, for example,
Fig. \ref{fig3}(b)), which connects the two Weyl nodes in the bulk
spectrum. To confirm that the surface states are indeed zero-energy
Majorana modes, in Fig. \ref{fig6}(d) we plot the wave-function of
one of the surface states at $k_{y}=k_{z}=0$. The wave-function is
well localized at the two boundaries and satisfies the desired symmetry
$u_{\sigma}(x)=e^{i\vartheta}v_{\sigma}^{*}(x)$ ($\sigma=\uparrow,\downarrow$)
for Majorana modes. It is interesting that with an in-plane Zeeman
field, the topological superfluid may support \emph{unidirectional}
chiral surface states, as demonstrated in Figs. \ref{fig6}(a) and
\ref{fig6}(b), which propagate in the same direction at the opposite
boundaries with the same velocity $v=\partial E/\partial k$ \cite{Hu2014}.
As discussed in Ref. \cite{Hu2014}, such unidirectional surface states
are actually a smoking-gun feature of the gapless topological superfluid.
Similar to the Majorana flat band, the unidirectional chiral surface
states also connect some points in the bulk spectrum, at which the
particle and hole branches touch with each other and the local spectrum
is approximately linear.

Our results of in-plane Zeeman field induced gapless topological superfluids
are not captured by a previous study based on the same model Hamiltonian
\cite{Xu2014}. In that work, in the absence of an out-of-plane Zeeman
field, the system was thought to always be in the gapless Fulde-Ferrell
superfluid \cite{Xu2014}. This discrepancy may originate from a different
definition of topological phase for the 2D slices, in the view of
dimensional reduction. On the other hand, we are not able to produce
the contour plot of the zero-energy quasiparticle spectrum at a large
in-plane Zeeman field, as shown by the blue line in Fig. 2(b) of Ref.
\cite{Xu2014}, which has \emph{connected} closed lines for nodal
points and therefore was thought topologically trivial. The contour
plots in our Figs. \ref{fig5}(c) and \ref{fig5}(d) are rather similar
to the one with a small out-of-plane Zeeman field as shown by the
red line in Fig. 2(b) of Ref. \cite{Xu2014}, which was also regarded
by those authors as topologically non-trivial.

\subsubsection{Equal in-plane and out-of-plane Zeeman fields}

\begin{figure}
\begin{centering}
\includegraphics[clip,width=0.48\textwidth]{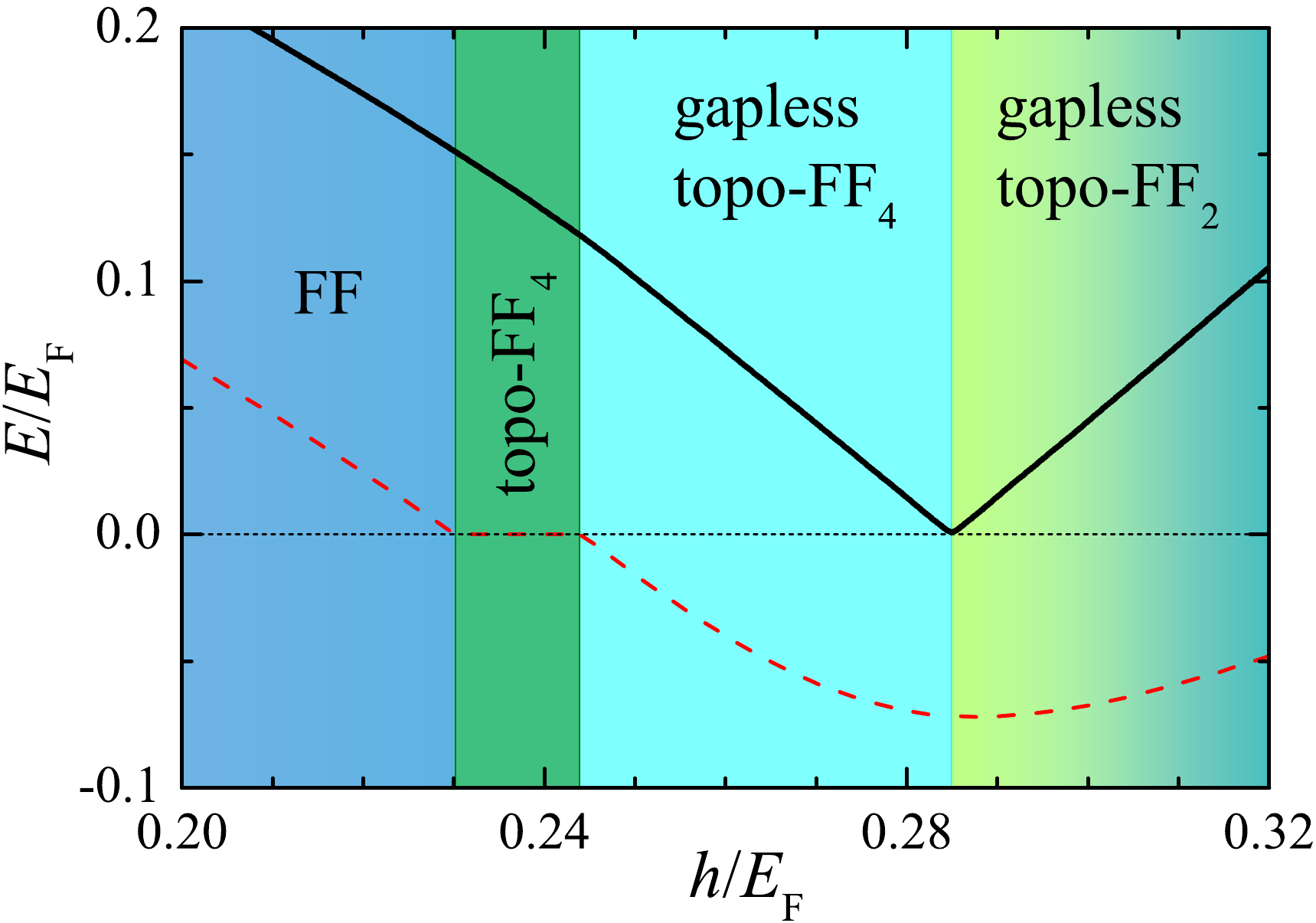} 
\par\end{centering}

\protect\caption{(Color online). (a) Evolution of the superfluid phase with equally
increasing the in-plane and out-of-plane Zeeman fields (i.e., $h_{\textrm{in}}=h_{\textrm{out}}=h$).
The system evolves from a gapped FF superfluid to a gapped topologically
nontrivial topo-FF$_{4}$ state, and then to a gapless topologically
non-trivial FF superfluid with either four (gapless topo-FF$_{4}$)
or two Weyl nodes/loops (gapless topo-FF$_{2}$). The solid and dashed
lines show the energy at $\mathbf{k}=0$ {[}$E_{2+}(\mathbf{k}=0)${]}
and the minimum energy {[}$\min E_{2+}(\mathbf{k})${]} of the lower
particle branch.}

\label{fig7} 
\end{figure}

\begin{figure}
\begin{centering}
\includegraphics[clip,width=0.48\textwidth]{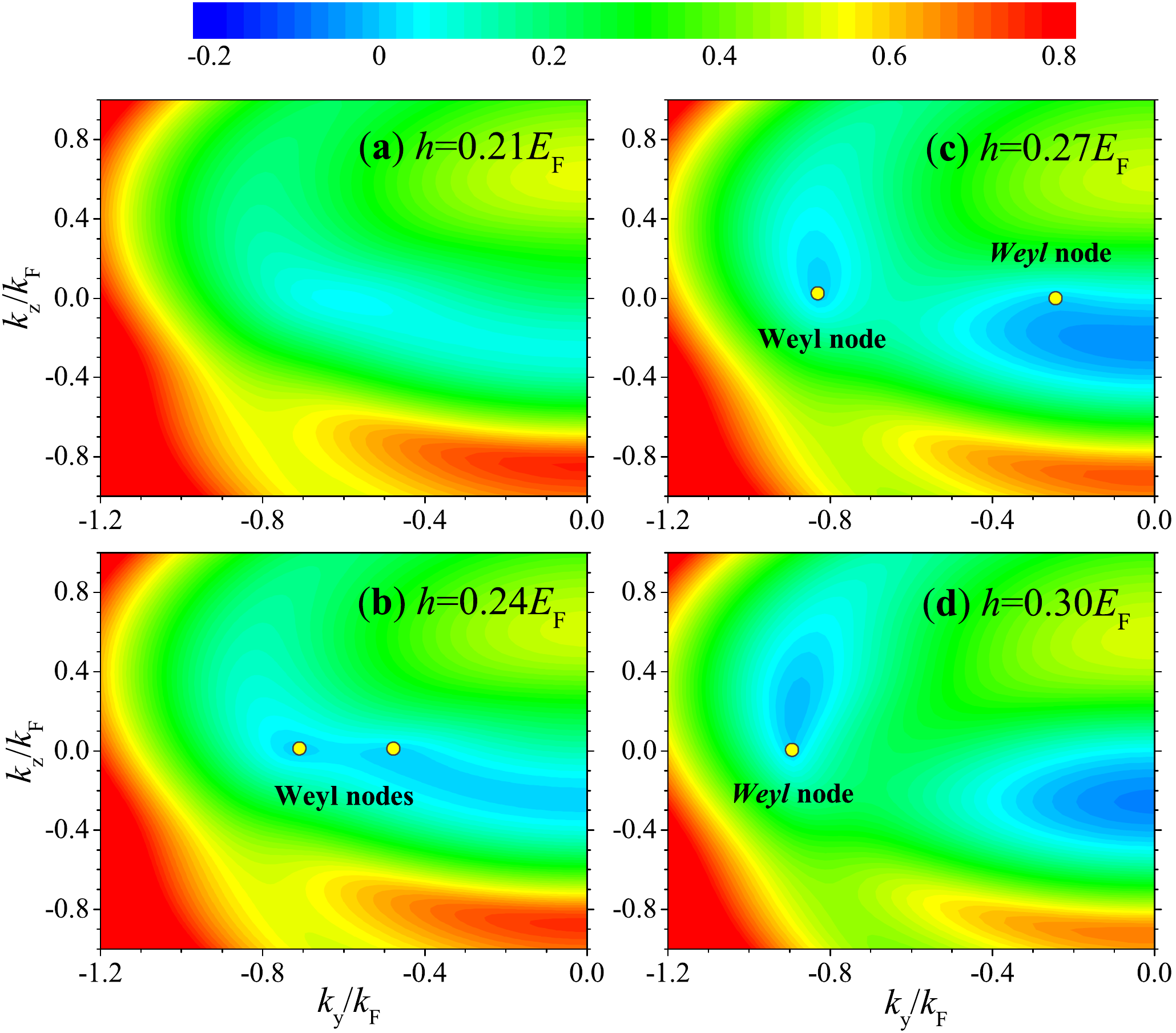} 
\par\end{centering}

\protect\caption{(Color online). Contour plots of the particle branch $E_{2+}(\mathbf{k})$
in the gapped FF phase (a), the gapped topo-FF$_{4}$ phase (b), the
gapless topo-FF$_{4}$ phase (c), and the gapless topo-FF$_{2}$ phase
(d). The Weyl nodes are highlighted by yellow solid circles. We note
that the in-plane and out-of-plane Zeeman fields are equal, $h_{\textrm{in}}=h_{\textrm{out}}=h$.
The spectrum is symmetric with respect to $k_{y}$, so we plot only
the left part with $k_{y}\leq0$.}

\label{fig8} 
\end{figure}
We now consider another special case with equal in-plane and out-of-plane
Zeeman fields. Figure \ref{fig7} shows the phase diagram determined
by monitoring the minimum energy and the local energy at $\mathbf{k}=0$
of the lower particle branch. The contour plots of the energy spectrum
at $k_{x}=0$ in different superfluid phases are reported in color
in Fig. \ref{fig8}. A gapped topological Fulde-Ferrell phase can
be easily identified from the zero-energy plateau in the minimum energy.
A close look at the energy spectrum (Fig. \ref{fig8}(b)) indicates
that this phase has four Weyl nodes and therefore should be classified
as a topo-FF$_{4}$ phase. With increasing the Zeeman field, nodal
points develop in the energy spectrum and the system enters the gapless
topological phase with four (Fig. \ref{fig8}(c)) and then two Weyl
nodes (Fig. \ref{fig8}(d)).

\begin{figure}
\begin{centering}
\includegraphics[clip,width=0.48\textwidth]{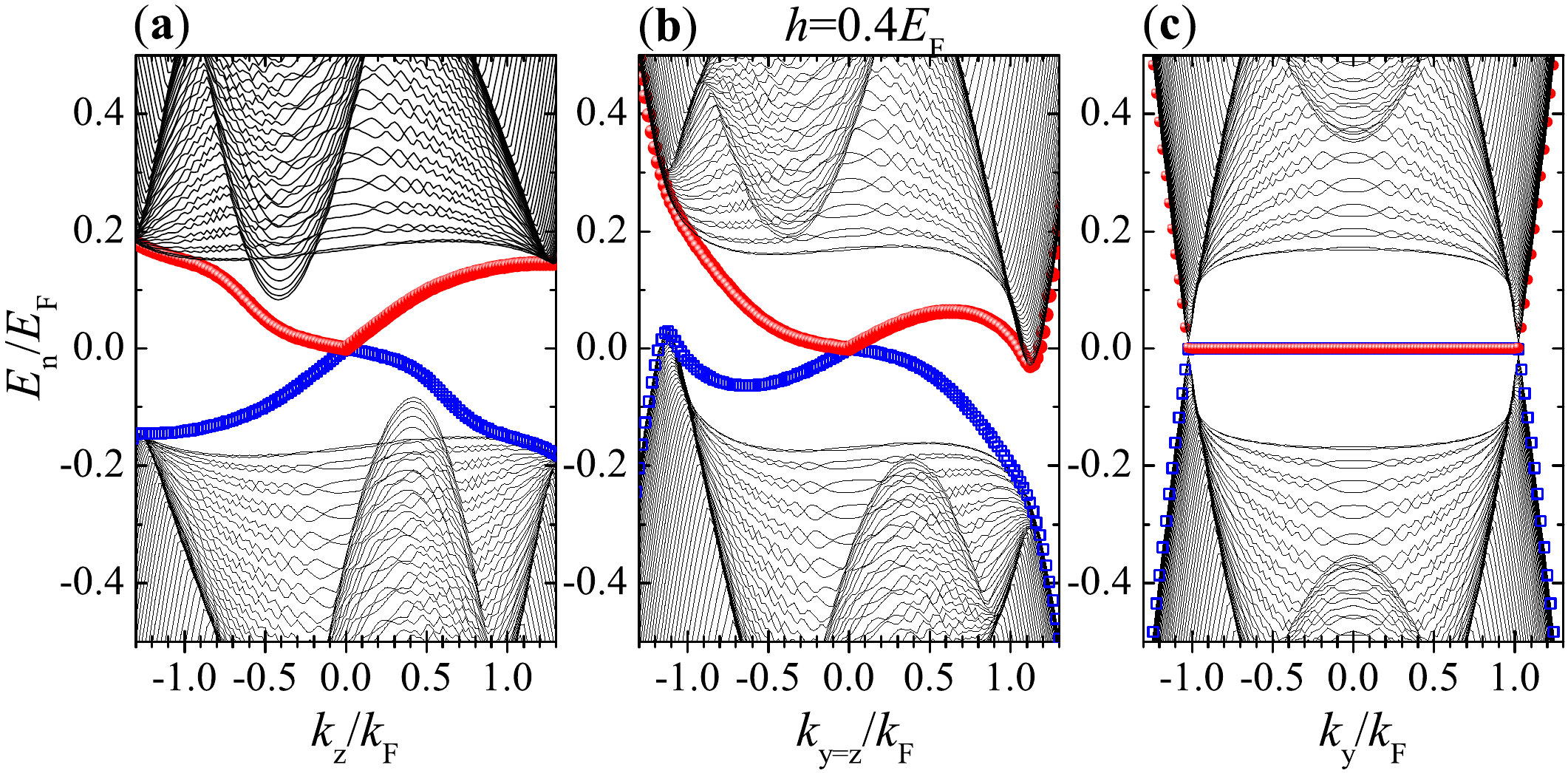} 
\par\end{centering}

\protect\caption{(Color online). Chiral surface modes with a hard-wall confinement
along the \textit{x}-direction in the gapless topo-FF$_{2}$ phase
($h_{\textrm{in}}=h_{\textrm{out}}=0.4E_{F}$), at $k_{y}=0$ (a),
along the $k_{y}=k_{z}$ line (b), or at $k_{z}=0$ (c).}

\label{fig9} 
\end{figure}

In Fig. \ref{fig9}, we show the chiral surface modes in the gapless
topo-FF$_{2}$ phase with $h_{\textrm{in}}=h_{\textrm{out}}=0.4E_{F}$.
Along the cut $k_{z}=0$ as shown in Fig. \ref{fig9}(c), we again
find the Majorana flat band as a function of the parameterized momentum
$k_{y}$. This is indeed a very robust feature of our system as the
Hamiltonian is dispersionless along the $k_{y}$-axis. When we make
cuts along other directions, i.e., along the $k_{y}=0$ and $k_{y}=k_{z}$
axis in Figs. \ref{fig9}(a) and \ref{fig9}(b), the surface states
do have dispersion. They propagate along the opposite direction at
the two boundaries, as a result of the \emph{gapped} energy spectrum
along the cuts \cite{Hu2014}, although the topological phase itself
is globally gapless in the bulk.

\subsubsection{Phase diagram on the $h_{\textrm{in}}$-$h_{\textrm{out}}$ plane}

By tuning in-plane and out-of-plane Zeeman fields, we obtain the zero-temperature
phase diagram, as displayed earlier in Fig. \ref{fig1}. The solid
line, which separates a usual Fulde-Ferrell superfluid from a topologically
non-trivial Fulde-Ferrell superfluid, is determined by the first appearance
of Weyl nodes in the energy of the lower particle branch along the
$k_{y}$-direction, $E_{2+}(k_{x}=0,k_{y},k_{z}=0)$ (see, for example,
Fig. \ref{fig4}(b)). The transition within different topological
phases (i.e., from four to two Weyl nodes), shown by the dot-dashed
line, can be conveniently determined by the closing of the local energy
gap at origin, i.e., $E_{2+}(\mathbf{k=0})=0$. On the other hand,
the gapless transition indicated by red solid circles in the figure
could be obtained from the critical field at which the minimum energy
$\min E_{2+}(\mathbf{k})$ becomes negative. It is worth noting that
the red solid circles are not smoothly connected. Physically, on the
left-hand side of these solid circles, the topological phase has occasionally
touched Weyl nodes along the $k_{y}$-axis, while on their right-hand
side, each Weyl node extends to form a loop of nodal points on the
$k_{y}$-$k_{z}$ plane, as we already discussed in Figs. \ref{fig5}(c)
and \ref{fig5}(d).

It is impressive that on the BCS side the topological Fulde-Ferrell
phases, either gapped or gapless, dominate in the phase diagram. In
particular, the gapless topological Fulde-Ferrell superfluid with
two Weyl nodes is already energetically favorable at a moderately
large in-plane Zeeman field (i.e., $h_{\textrm{in}}>0.35E_{F}$).

\subsection{Finite-temperature phase diagram}

\begin{figure}
\begin{centering}
\includegraphics[clip,width=0.48\textwidth]{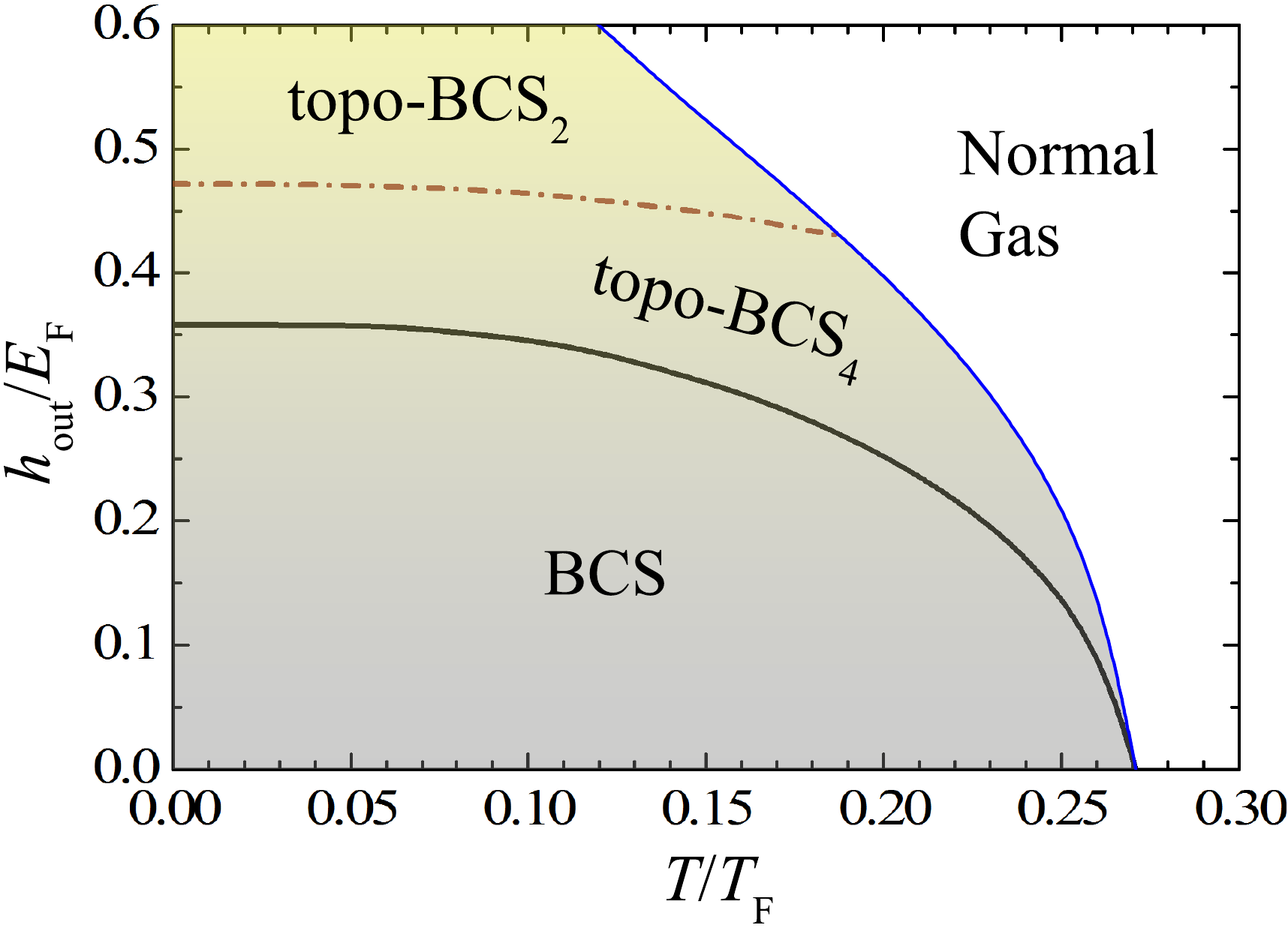} 
\par\end{centering}

\protect\caption{(Color online). Finite temperature phase diagram on the $T$-$h_{\textrm{out}}$
plane at zero in-plane Zeeman field $h_{\textrm{in}}=0$.}

\label{fig10} 
\end{figure}

\begin{figure}
\begin{centering}
\includegraphics[clip,width=0.48\textwidth]{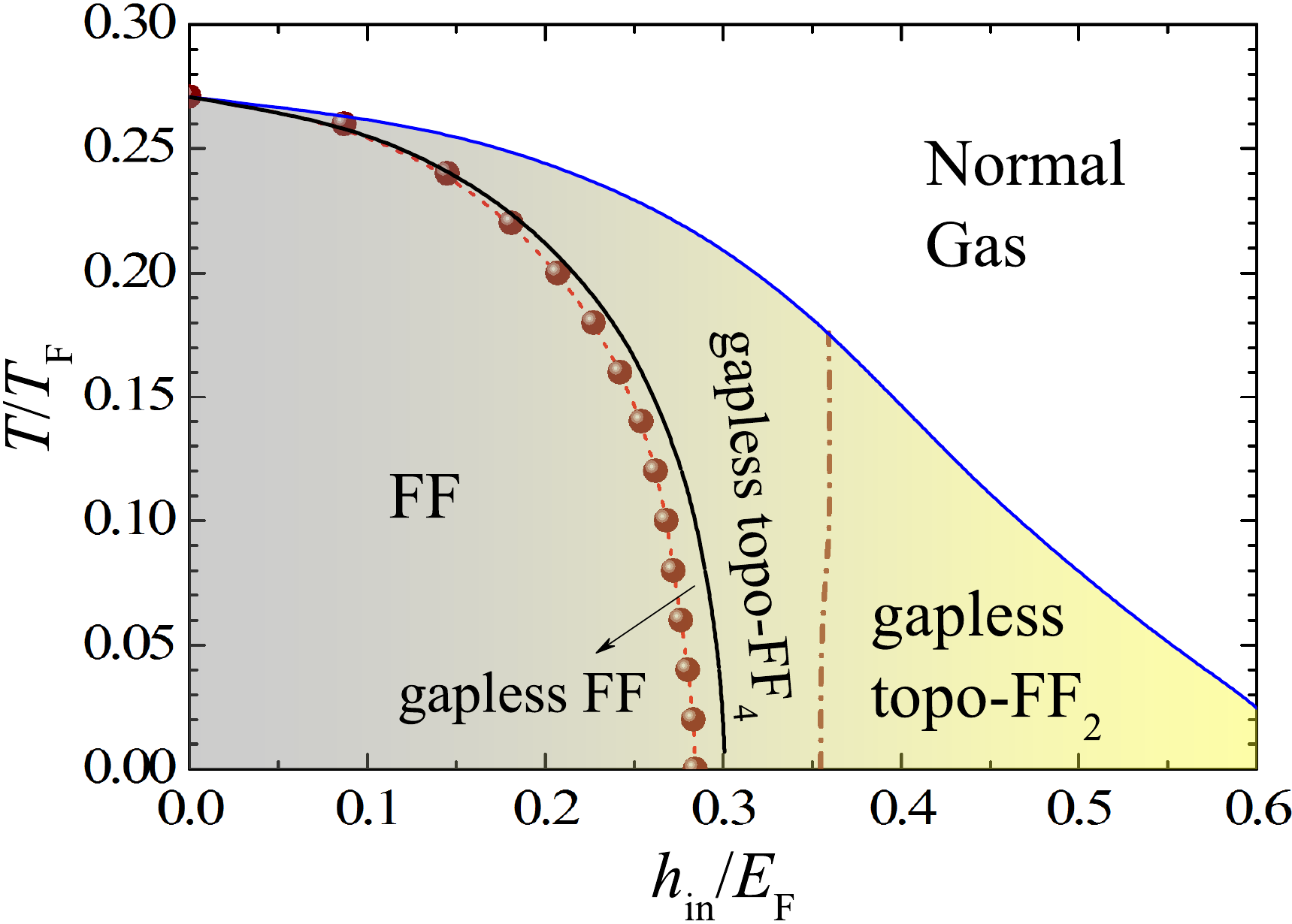} 
\par\end{centering}

\protect\caption{(Color online). Finite temperature phase diagram on the $h_{\textrm{in}}$-$T$
plane at zero out-of-plane Zeeman field $h_{\textrm{out}}=0$.}

\label{fig11} 
\end{figure}

\begin{figure}
\begin{centering}
\includegraphics[clip,width=0.48\textwidth]{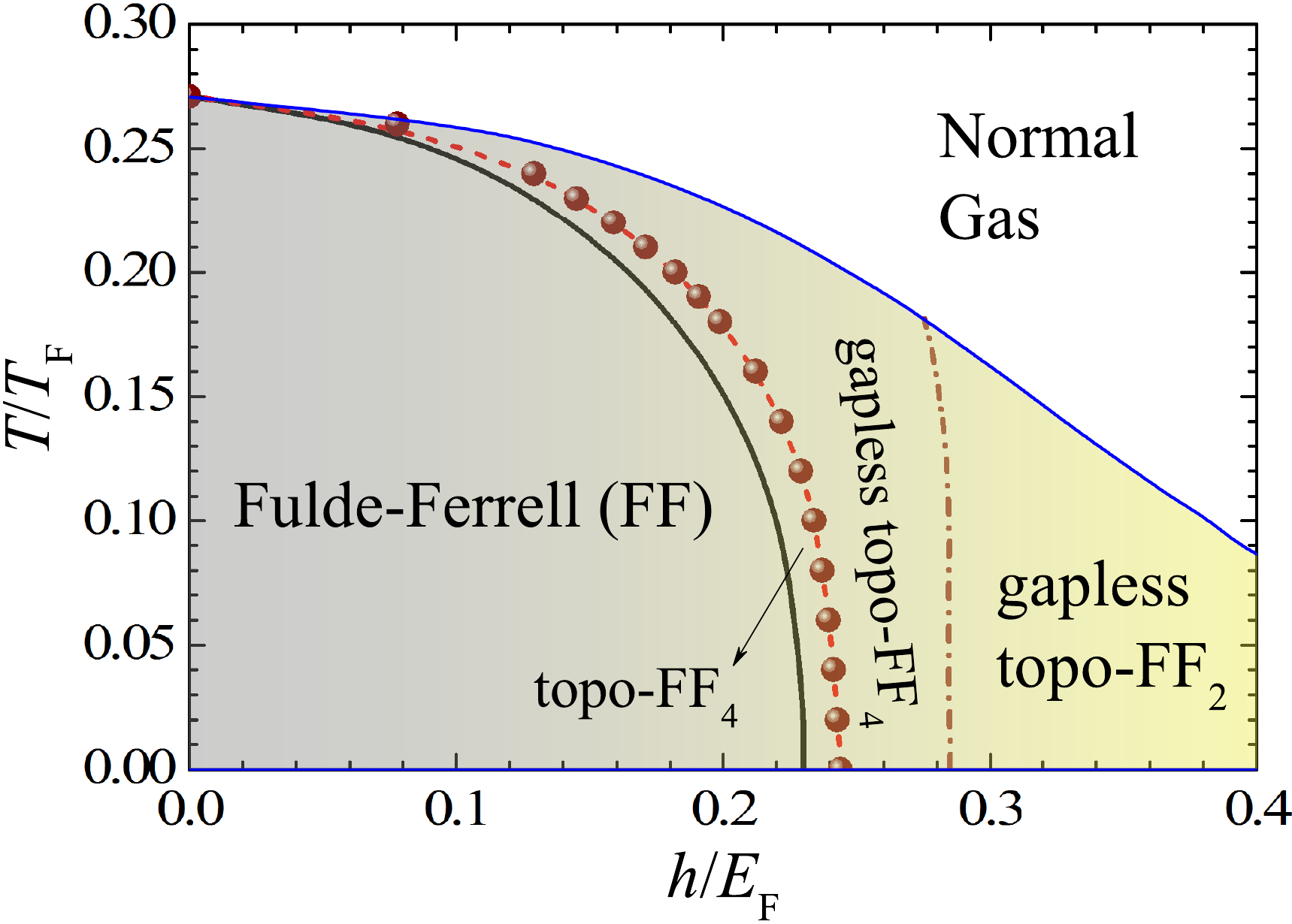} 
\par\end{centering}

\protect\caption{(Color online). Finite temperature phase diagram on the $h$-$T$
plane with equal in-plane and out-of-plane Zeeman fields $h_{\textrm{in}}=h_{\textrm{out}}=h$.}

\label{fig12} 
\end{figure}

In current cold-atom laboratories, the lowest accessible experimental
temperature is about $0.05-0.1T_{F}$ \cite{Ku2012}. It is thus useful
to consider how the different superfluid phases involve as temperature
increases. In Figs. \ref{fig10}, \ref{fig11} and \ref{fig12}, we
show the finite-temperature phase diagrams at different configurations
of Zeeman fields. Our mean-field theory is less reliable at finite
temperature. However, on the BCS side with a weak interaction parameter
$1/(k_{F}a_{s})=-0.5$, we anticipate that it will provide a very
good semi-quantitative description.

In the presence of out-of-plane Zeeman field (Fig. \ref{fig10}),
the finite-temperature phase diagram has been earlier discussed by
Seo and co-workers \cite{Seo2013}. It contains two topological BCS
phases and is relatively easy to understand. In the other two cases
with in-plane Zeeman field only (Fig. \ref{fig11}) or equal in-plane
and out-of-plane Zeeman fields (Fig. \ref{fig12}), gapless topological
Fulde-Ferrell superfluids emerge at sufficiently large field strength.
It is remarkable that the phase space for these gapless topological
superfluids is very significant at finite temperature. In particular,
when the system is cooled down from a normal state at high temperature,
the pairing instability first occurs towards the formation of gapless
topological phases. The typical superfluid transition temperature
is about $0.15T_{F}$, which is clearly within the reach in current
cold-atom experiments \cite{Ku2012}.

\subsection{Phase diagram with imperfect Rashba spin-orbit coupling }

\begin{figure}
\begin{centering}
\includegraphics[clip,width=0.48\textwidth]{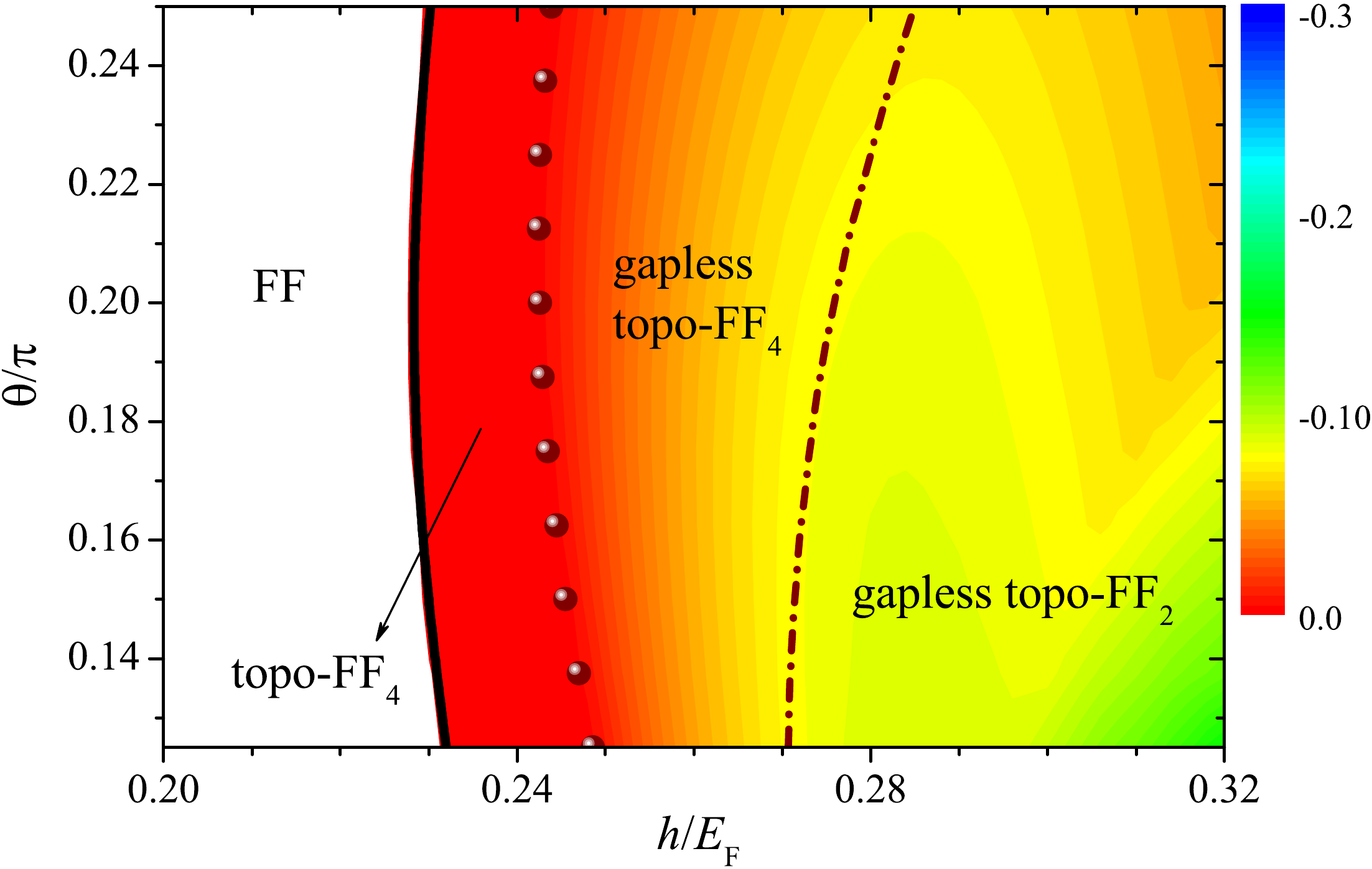} 
\par\end{centering}

\protect\caption{(Color online). Zero-temperature phase diagram with an anisotropic
spin-orbit coupling $\lambda(\cos\theta\hat{k}_{z}\sigma_{z}+\sin\theta\hat{k}_{x}\sigma_{x})$
and equal in-plane and out-of-plane Zeeman fields $h_{\textrm{in}}=h_{\textrm{out}}=h$.
The Rashba spin-orbit coupling corresponds to the case with $\theta=\pi/4$.
The color map shows the lowest energy in the particle branches, in
units of $E_{F}$. Here, we use a spin-orbit coupling strength $\lambda=E_{F}/k_{F}$
and an interaction strength $1/(k_{F}a_{s})=-0.5$. }

\label{fig13} 
\end{figure}

\begin{figure}
\begin{centering}
\includegraphics[clip,width=0.48\textwidth]{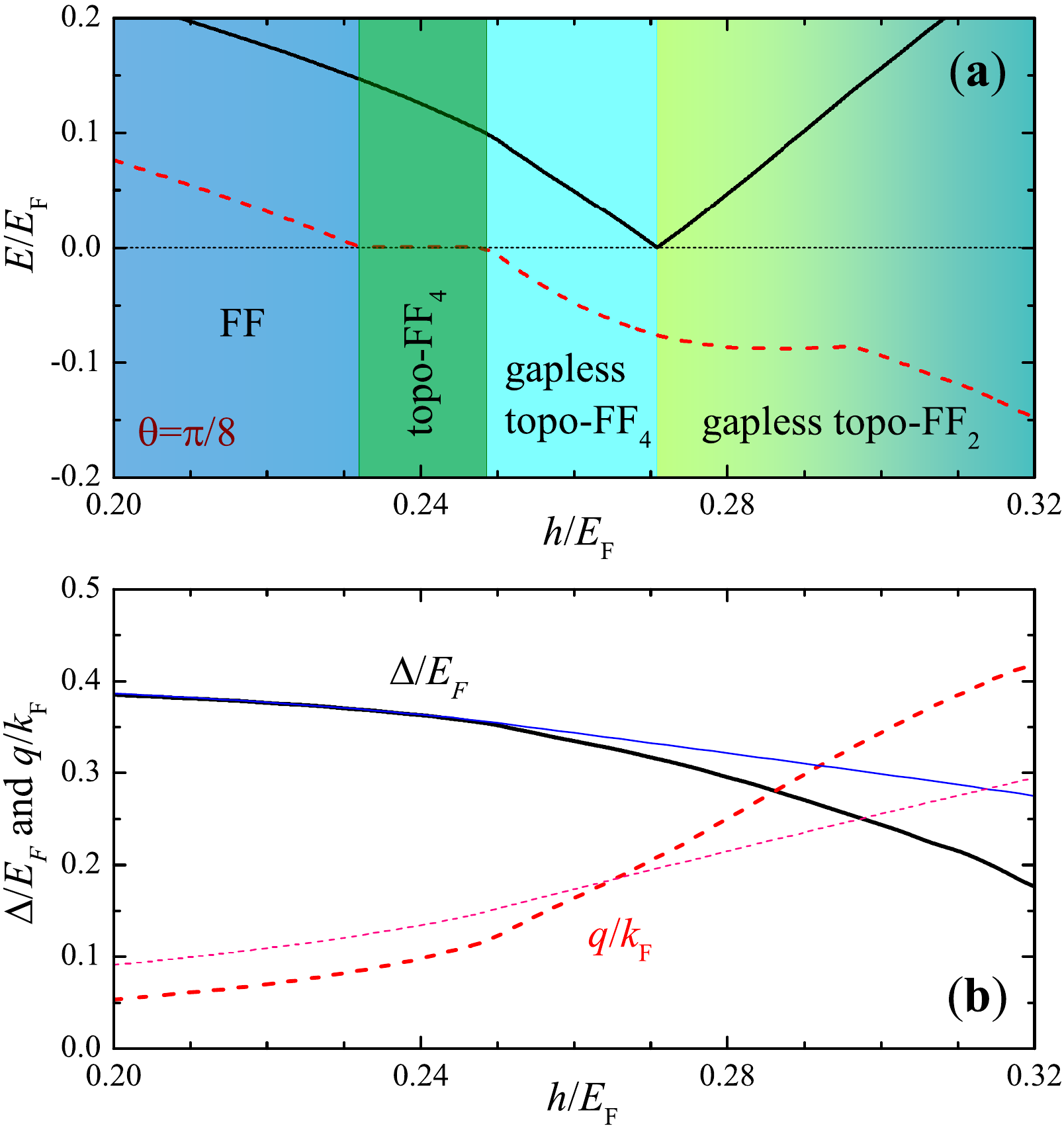} 
\par\end{centering}

\protect\caption{(Color online). (a) Evolution of the superfluid phase with equally
increasing the in-plane and out-plane Zeeman fields, and with an imperfect
Rashba spin-orbit coupling at the angle $\theta=\pi/8$. The system
evolves from a gapped FF superfluid to a gapped topologically nontrivial
topo-FF$_{4}$ state, and then to a gapless topologically non-trivial
FF superfluid with either four (gapless topo-FF$_{4}$) or two Weyl
nodes/loops (gapless topo-FF$_{2}$). The solid and dashed lines show
the energy at $\mathbf{k}=0$ {[}$E_{2+}(\mathbf{k}=0)${]} and the
minimum energy {[}$\min E_{2+}(\mathbf{k})${]} of the lower particle
branch. (b) The pairing gap (solid line) and FF pairing momentum (dashed
line) as a function of the Zeeman field. The thick and thin lines
correspond to the cases with $\theta=\pi/8$ and $\theta=\pi/4$,
respectively.}

\label{fig14} 
\end{figure}

We now turn to consider a general form of spin-orbit coupling with
$\lambda_{z}=\lambda\cos\theta$ and $\lambda_{x}=\lambda\sin\theta$,
away from the Rashba spin-orbit coupling that corresponds to $\theta=\pi/4$.
This investigation is helpful for future experiments, as the synthetic
spin-orbit coupling - to be realized in cold-atom laboratories - may
not acquire the perfect form of Rashba spin-orbit coupling. In Fig.
\ref{fig13}, we show the evolution of the zero-temperature phase
diagram as a function of the angle $\theta$ down to $\theta=\pi/8$,
in the presence of equal in-plane and out-of-plane Zeeman fields.
The Zeeman field dependence of the minimum energy $\min E_{2+}(\mathbf{k})$
and the local energy at $\mathbf{k}=0$ $E_{2+}(\mathbf{k}=0)$ of
the lower particle branch in the case of $\theta=\pi/8$, where the
component of Dresselhaus spin-orbit coupling becomes significant,
are reported in Fig. \ref{fig14}(a).

The phase diagram is less affected by tuning the form of spin-orbit
coupling. With decreasing $\theta$, the phase space for the topological
phases (including gapped or gapless topo-FF$_{4}$, and gapless topo-FF$_{2}$)
is essentially unchanged. It is remarkable that the most interesting
gapless topo-FF$_{2}$ phase seems to be more favorable away from
the Rashba spin-orbit coupling. This could be attributed to the increasing
Fulde-Ferrell momentum $q$ at large Zeeman field, as shown in Fig.
\ref{fig14}(b), which favors the gapless phase and also the topological
phase with two Weyl nodes.

\section{Conclusions}

In conclusions, we have presented a systematic investigation of the
superfluid phases of a spin-orbit coupled Fermi gas with both in-plane
and out-of-plane Zeeman fields, by using a dimensional reduction picture
\cite{Sau2012}. Driven by large Zeeman fields, in particular, the
large in-plane Zeeman field, the system supports a number of intriguing
features including Fulde-Ferrell pairing, topological superfluids
with Majorana fermions in real space and Weyl fermions in momentum
space, and also novel gapless topological superfluids. We have considered
the phase diagram at both zero temperature and finite temperature.
We have mainly focused on Rashba spin-orbit coupling. The case with
an imperfect Rashba spin-orbit coupling has also been addressed, from
a realistic experimental point of view. Our work complements the earlier
investigation by Xu and co-workers \cite{Xu2014}, by providing a
richer phase diagram at zero temperature and extended phase diagrams
at finite temperature and with more general spin-orbit coupling.

In future studies, it will be useful to consider the pair fluctuations
which become very significant at the BEC-BCS crossover at both zero
and finite temperatures. The finite-temperature pair fluctuations
for our system have been recently addressed within a pseudogap theory
\cite{Zheng2014}. More rigorous treatments could rely on the different
many-body $T$-matrix theories within the ladder approximation \cite{Hu2006,Hu2008}.
\begin{acknowledgments}
X.-J.L and H.H. were supported by the ARC Discovery Projects (Grant
Nos. FT140100003, FT130100815, DP140103231 and DP140100637) and the
National Key Basic Research Special Foundation of China (NKBRSFC-China)
(Grant No. 2011CB921502). HP was supported by the NSF, the Welch Foundation
(Grant No. C-1669) and the DARPA OLE program.\end{acknowledgments}


\begin{thebibliography}{10}
\bibitem{footnote1}This is a branch cut at $k_{z}=0$ to a 3D surface
plot of the energy spectrum, as shown, for example, in Figs. \ref{fig2}(b)
and \ref{fig2}(c).
\end{thebibliography}
\end{document}